# Highly fluorescent copper nanoclusters for sensing and bioimaging

Yu An [a,#], Ying Ren [b,#], Jing Tang [d,*], Jun Chen [c,*] and Baisong Chang [a,*]

[a] State Key Laboratory of Advanced Technology for Materials Synthesis and Processing, Wuhan University of Technology, Wuhan 430070, P.R. China

[b] Department of Radiology, Shengjing Hospital of China Medical University, Shenyang, 110004, P.R. China

[c] Department of Bioengineering, University of California, Los Angeles, Los Angeles, California 90095, USA

[d] Department of Materials Science and Engineering, Stanford University, Stanford, California 94306, USA

\* Correspondence: jingtang@stanford.edu (J.T.), chang@whut.edu.cn (B.C.), jun.chen@ucla.edu (J.C.)

[#] These authors contributed equally to this work.

**Abstract:** Metal nanoclusters (NCs), typically consisting of a few to tens of metal atoms, bridge the gap between organometallic compounds and crystalline metal nanoparticles. As their size approaches the Fermi wavelength of electrons, metal NCs exhibit discrete energy levels, which in turn results in the emergence of intriguing physical and chemical (or physicochemical) properties, especially strong fluorescence. In the past few decades, dramatic growth has been witnessed in the development of different types of noble metal NCs (mainly AuNCs and AgNCs). However, compared with noble metals, copper is a relatively earth-abundant and cost-effective metal. Theoretical and experimental studies have shown that copper NCs (CuNCs) possess unique catalytic and photoluminescent properties. In this context, CuNCs are emerging as a new class of nontoxic, economic, and effective phosphors and catalysts, drawing significant interest across the life and medical sciences. To highlight these achievements, this review begins by providing an overview of a multitude of factors that play central roles in the fluorescence of CuNCs. Additionally, a critical perspective of how the aggregation of CuNCs can efficiently improve the florescent stability, tunability, and intensity is also discussed. Following, we present representative applications of CuNCs in detection and bioimaging. Finally, we outline current challenges and our perspective on the development of CuNCs.

## 1. Introduction

When decreasing the size of nanoparticles so that it approaches the Fermi wavelength of an electron, novel optical, electrical and magnetic properties appear (Deng et al. 2018a; Li et al. 2016b; Moghadam and Rahaie 2019; Wang et al. 2018c). Commonly termed nanoclusters (NCs), these ultra-small nanoparticles, bridging the missing link between atoms and nanocrystals, have attracted considerable attention in both fundamental research and practical applications (Bagheri et al. 2017; Basu et al. 2019; Sun et al. 2019; Wang et al. 2016b). Benefiting from the great progress in nanosynthetic chemistry, high-quality Copper NCs (CuNCs) with tailored size and good stability can be easily obtained. CuNCs, therefore, constitute an active research direction due to their unique properties, such as high luminous efficiency, long fluorescence lifetime, good optical and chemical stability, and large stokes displacement, which distinguish them as a new type of fluorescent probe for optical sensing and bioimaging/labeling (Cao et al. 2014; Wang et al. 2017).

In this review, we highlight recent developments of CuNCs, with a particular emphasis on the multiple factors affecting fluorescent properties of CuNCs, as well as the aggregation-induced emission enhancement (AIEE) effect. To appreciate these advances, applications focused on the detection of ions, macromolecules and bioimaging are summarized. Lastly, we will provide a brief conclusion, and will address future challenges in the rapidly growing field of fluorescent CuNCs.

## 2. Multiple factors that govern the fluorescence of CuNCs

Due to the size of CuNCs falls within the sub-nanometer regime, delicate control of experimental conditions is critical for the synthesis of high-quality CuNCs. A series of factors, such as core size, surface ligands, molecular structure, charge, ionic strength, pH and temperature, have been documented to play a significant effect on the fluorescence of CuNCs.

### *2.1. Size-dependent fluorescence and ligand effect*

As metal size decreases, especially approaching the Fermi wavelength of electrons, CuNCs generate strong photoluminescent properties (Qian et al. 2012). The fluorescence of CuNCs originates from electronic transitions between occupied d bands and the aforementioned Fermi level (ca. sp bands), that is to say, the electronic transitions between the highest occupied orbital and the lowest unoccupied orbital (HOMO-LUMO) (Lu and Chen 2012). In 1969, Mooradian et al. (Mooradian 1969) first observed the photoluminescence of copper metal. The photoluminescence was further attributed to the surface plasmon oscillations in copper nanostructures, the electromagnetic enhancement of electrons in the sp conduction band below the Fermi level, and the vacancy in the d-band hole (Darugar et al. 2006).

Size dependence of the optical properties of CuNCs has been reported both theoretically and experimentally. The results typically follow the Vazquez-Vazquez's theory (Vazquez-Vazquez et al. 2009), which explains the fluorescence of CuNCs at the atomic level. More importantly, the authors presented evidence for how the size of CuNCs can be controlled by adjusting the percentage of reducing agent (α) used for synthesis without the use of thiolates. As shown in Fig. 1, when the proportion of reducing agent was below 10%, $Cu_n$ clusters with a small number of atoms were obtained. Specifically, when n was smaller than 10, CuNCs with high photoluminescence were observed. As the proportion of reducing agent increased, the luminescent $Cu_n$ clusters disappeared, resulting in the appearance of clusters with a larger particle size accompanied by a red shift of the UV-visible absorption band.

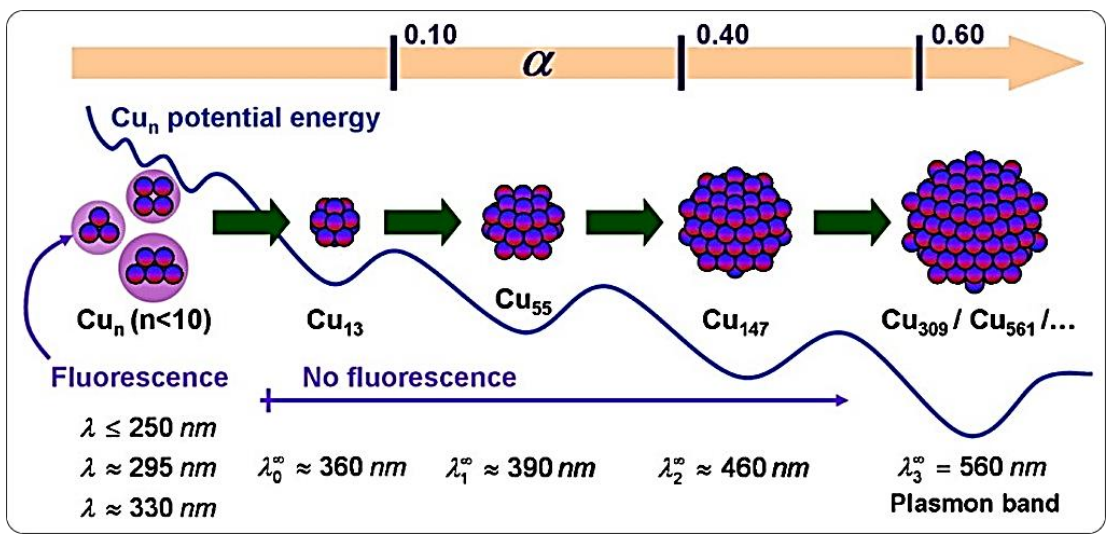

**Fig. 1.** The evolution of the copper cluster size with increasing reducing agent (α) values. When the proportion of reducing agent used was below 10%, $Cu_n$ clusters with a small number of atoms were obtained. Especially when n was smaller than 10, CuNCs with high photoluminescence were observed. Estimated wavelengths for each cluster were shown at the bottom. The potential energy trend was schematically drawn for several $Cu_n$ clusters (reprinted from (Vazquez-Vazquez et al. 2009) with permission from American Chemical Society).

Over the past few years, assembling relatively unstable CuNCs into uniformly larger nanostructures with controllable size and shape has provided new opportunities for applications of CuNCs, ranging from electronics, and optics to nanomedicine (Nie et al. 2010; Tan et al. 2017). The use of self-assembly strategies to enhance fluorescence emission of nano-components has attracted much attention, such as with the successful synthesis of copper nanosheets and nanoribbons (Wu et al. 2015). Unfortunately, most self-assembled components were synthesized in organic solvents, which not only limited their biological applications, but also hindered their use in optical devices. With this in mind, it would be of great significance to successfully achieve self-assembly of copper nanomaterials in an aqueous solution. Inspired by the microemulsion method, amphiphilic compounds offer a novel strategy. The amphiphilic block copolymer in solution provides a unique environment for controlling the growth of NPs and building a uniform assembled material, and it can further facilitate structural self-assembly of controlled size and shape (Elsabahy et al. 2015; Mai and Eisenberg 2012).

Very recently, in 2019, Zhou and colleagues reported a practical and robust strategy for *in situ* synthesis of highly luminescent Cu nanoassemblies with uniform morphology and remarkable stability (Zhou et al. 2019). In their study, three types of triblock copolymers, pluronic F68, F108 and F127, were used, each having different numbers of hydrophilic and hydrophobic blocks. The Cu nanoassemblies were synthesized using a multidentate thiol dipentaerythritol hexakis (S6) as the ligand and binder (Fig. 2a). Its rigid structure not only bridged CuNCs, but also induced enhanced component emission, and ensured photophysical and structural stability of the component in a physiological environment. The hydrophilic group on the surface of pluronic enabled the synthesized Cu nanoassemblies to dissolve in water. The oxidation of the metal core of the CuNCs component could also be prevented due to the unique core-shell structure.

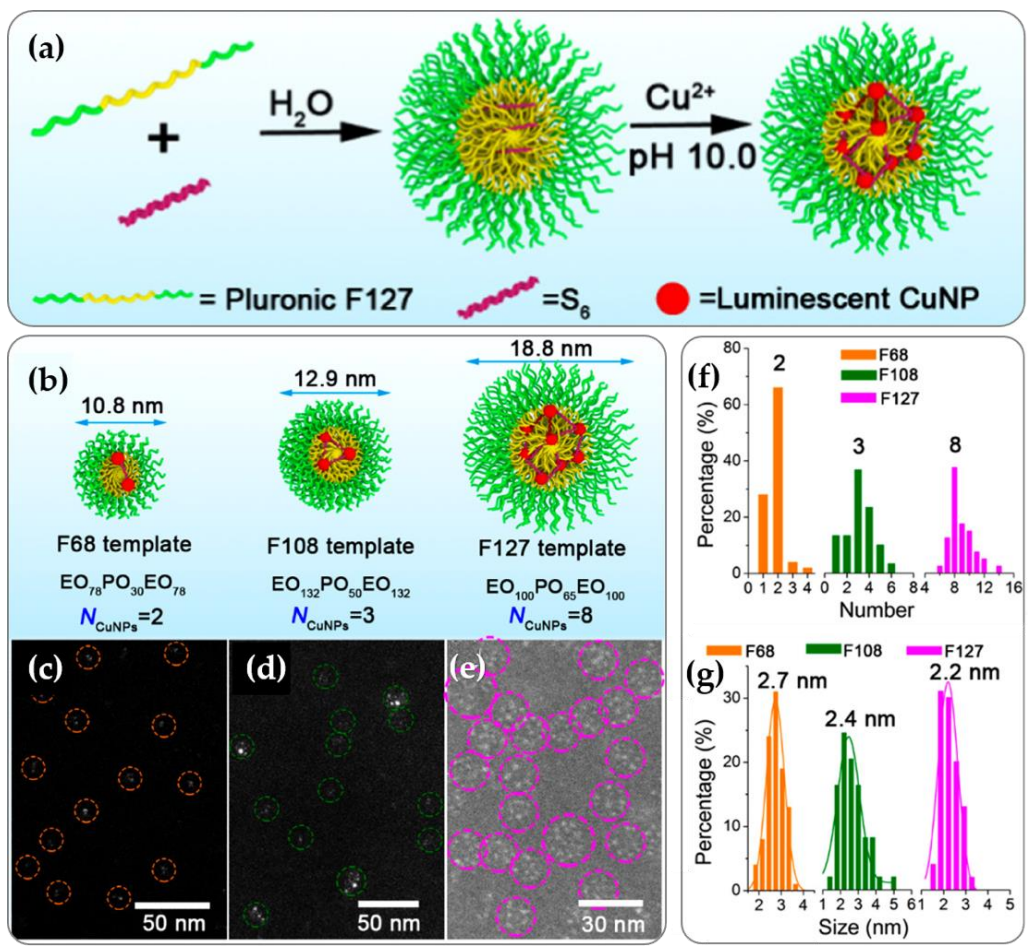

**Fig. 2.** (a) Schematic diagram of the synthesis of luminescent CuNPs. (b) Schematic illustration of the Cu nanoassemblies with controllable encapsulated CuNP number through varying block segments of the template. The High-Angle Annular Dark Field scanning transmission electron microscopy (HAADF-STEM) images of the CuNP assemblies using pluronic F68 (c), F108 (d), and F127 (e) as the template. (f) Number distribution of encapsulated CuNPs in an assembly and (g) size distribution of CuNP metal core in the assemblies (reprinted from (Zhou et al. 2019) with permission from American Chemical Society).

Furthermore, Cu nanoassembly growth was affected by the three different hydrophobic core sizes of the spherical micelle templates formed by the self-assembly of the triblock copolymers. As shown in Fig. 2b, when the template was changed from F68 to F108 and then F127, the dimensions of the Cu nanoassemblies changed from 10.8±1.2 nm, 12.9±1.4 nm to 18.8±2.4 nm, respectively. The main number of CuNPs encapsulated in the component went from about 2 to 3, and then to 8 (Fig. 2c-f). In the hydrophobic core, the three types of Cu nanoassemblies consisted of multiple homogeneously crosslinked spherical metal cores with an average size of 2.2-2.7 nm (Fig. 2g). Generally, the proposed strategy provided a good platform for realizing the synthesis of controllably packaged Cu nanoassemblies. As the amount of mercaptans in the ligand increased, the emission of the CuNPs module was significantly enhanced. The cross-linking between the ligand rigidity and ultra-small CuNPs resulted in enhanced fluorescence. The two properties interacted, and the major number of encapsulated CuNPs with a high quantum yield of 7.3% decreased from 11 to 8. In addition, the number of central atoms seemed to explain the enhancement of fluorescence intensity described above. The above results suggest that the integration of amphiphilic compounds with the self-assembly of CuNCs is a promising method to prepare stable, controlled and highly luminescent Cu nanoassemblies.

As anticipated, the type and state of surface ligands of CuNCs greatly influenced fluorescent properties. Taking AuNCs as an example, Wu et al. (Wu and Jin 2010) summarized the relationship between fluorescence emission of gold nanoclusters and ligands, and then gave three suggestions, which may also be applicable to CuNCs. In order to enhance the fluorescence emission of CuNCs, we should (1) increase the electron donation ability of ligands; (2) increase the electropositivity of the metal core, if the core of nanoparticles can maintain multiple charge states; and (3) use ligands with electron-rich atoms and groups, which may be the most effective strategy.

The function of the protein directly depends on its three-dimensional conformation (Moghadam and Rahaie 2019). It is well known that bovine serum albumin (BSA) has different conformations at different pH values, which will govern the physical properties of nanomaterials (Ghosh et al. 2012). The successful use of fluorescent CuNCs as a pH sensor was first reported in 2014 by Wang et al (Wang et al. 2014). It was found that the fluorescence intensity of BSA-CuNCs increased nearly 2-fold when the pH was lowered from 12 to 6, and these changes were reversible between pH 6-12. However, there was no significant change in emission positions at different pH values. In order to resolve this phenomenon, they studied the secondary structure of BSA in the reaction systems at different pH values by circular dichroism (CD) spectroscopy. It was found that, as the pH of the solution decreased, more alpha helices were stretched and converted into beta sheets, which may result in the damage or cleavage of hydrogen bonds. Accordingly, changes in fluorescence at different pH values could be attributed to the corresponding structural changes in BSA.

It should be noted that the luminescence mechanism is not only related to the quantum confinement effect of the metal core, but also the ligands on the surface of CuNCs (Chen et al. 2014; Lin et al. 2015; Luo et al. 2012). Coordination between ligands and the metal core usually results in ligand-metal charge transfer (LMCT) or ligand-metal-metal charge transfer (LMMCT), and radiation relaxation through the triplet state of metal center (Luo et al. 2012). Compared with metal centered electron transition induced fluorescence, LMCT or LMMCT induced emissions always show a longer fluorescence lifetime, which is partly because LMCT/LMMCT can control the excited state relaxation dynamics (Sharma and Dormidontova 2017).

## 2.2. Cu⋯Cu interantion

The Group IB metal in the +1 oxidation state generally has a very short metal-metal bond length. This unusual physicochemical property mainly originates from the interaction of $d^{10}$-$d^{10}$, which is also called closed-shell interactions (CSI). Although the strength of CSI is not comparable to the strength of covalent or ionic bonds, metallophilic attractions are related to correlation and relativistic effects, which have been regarded as important for explaining the luminescent behavior of metal compounds (Hermann et al. 2001). Although a majority of NCs are more likely to aggregate and oxidize as their size decreases, it is worth noting that moderate oxidation of the surface of NCs would ensure stability, and therefore enhance the fluorescence emission (Wang and Huang 2013). As the binding energy of Cu(0) is only 0.1 eV away from that of Cu(I) in X-ray photoelectron spectroscopy, it is not possible to exclude the formation of Cu(I). Therefore, the valence state of the obtained CuNCs most likely lies between 0 and +1. Wang et al. (Wang and Huang 2013) studied the effect of reducing agents on the fluorescent emission of thiol-protected CuNCs. They found that the addition of $NaBH_4$ rapidly decreased the fluorescence of as-prepared glutathione (GSH) capped CuNCs (GSH-CuNCs). In addition, a new broad peak appeared in the XRD spectrum that was consistent with the copper particles. It is known that the charge transfer from S to metal core has a predominant effect on fluorescent enhancement. The oxidation of CuNCs core can thus promote the charge transfer efficiency and further improve the fluorescent signals (Chen et al. 2015).

Since the solid-state emission heavily depends on the arrangement and molecular structure in the solid state, it is widely observed that mechanical grinding will produce a direct influence on the structural packing mode and, in turn, change the emission wavelength (Sagara et al. 2007). In 2014, Quentin et al. (Benito et al. 2014) provided new insights into the properties of mechanochromic luminescence through their in-depth study of luminescent copper iodide clusters. They found that the synthesized polycrystalline copper iodide clusters had the properties of mechanical discoloration and thermochromism, which caused a significant change in the color of luminescence when the solid sample was ground up. The initial green luminescence was converted to a yellow luminescence, while the powder color remained white. Grinding did not induce other chemical reactions, and the molecular structure of clusters stayed constant in the powders. After grinding, the product changed from crystalline to amorphous. By defining the Cu-iodine and Cu-phosphorus bond, the iodine-Cu-iodine bond angle indirectly determined the Cu-Cu distance. Thus, the offset of the emission spectrum correlated with the variation of Cu-Cu distance. Additionally, polymorphism and mechanical coloration can manipulate different thermochromic luminescence properties of clusters (Benito et al. 2014). The samples before and after grinding exhibited different emissions in liquid nitrogen. As the temperature increased, their emission returned to their original state.

The effect of Cu···Cu interaction in the liquid phase is similar to that described above. Wu et al. (Wu et al. 2015) also expressed a similar viewpoint through their self-assembly strategy of 1-dodecyl mercaptan (DT) terminated CuNCs. They declared that the compactness played an important role in the emission of NC components. High density not only enhanced the affinity interactions between NCs, but also greatly promoted the relaxation kinetics of excited states through radiation pathways (Jia et al. 2013; Luo et al. 2012); it strongly restricted the vibration and rotation of end-capped ligands, thus enhancing the emission intensity of self-assembly structures (Luo et al. 2012; Mei et al. 2014). In addition, the emission energy of the module depended on the distance between atoms. The improved compactness could increase the average distance between atoms, resulting in blue shift of emission from NC modules. In their research, NCs self-assembled through strong dipole-dipole interactions. Different fluorescence emissions were observed in self-assembled copper nanoribbons (Fig. 3a, b) and nanosheets (Fig. 3d, e). Although both of the nanocomponents were composed of a single NC rather than crystalline Cu, the nanosheets emitted blue fluorescence (Fig. 3c), while the nanoribbons emitted blue-green fluorescence (Fig. 3f). For nanosheets with loose NCs alignment, the cuprophilic interactions between NCs was weaker than that of nanoribbons. As a result, the average Cu(I)···Cu(I) distance was shorter than that of the nanoribbon, resulting in a decrease in emission energy and a red shift in the spectrum. Furthermore, they demonstrated that the emission of assembled nanomaterials was related to temperature. As for the nanoribbons, the intra- and inter-NCs Cu(I)···Cu(I) distances were simultaneously shortened under low temperature, resulting in red emission shifts (Fig. 3g). However, for the sheets, further temperature lowering made the increase in inter-NCs Cu(I)···Cu(I) distance dominant, resulting in a blue emission shift (Fig. 3h). Moreover, their group reported that metal defects contributed to emission red-shift (Wu et al. 2017). This discovery has then opened up new avenues for the application of copper nanomaterials in luminescent materials.

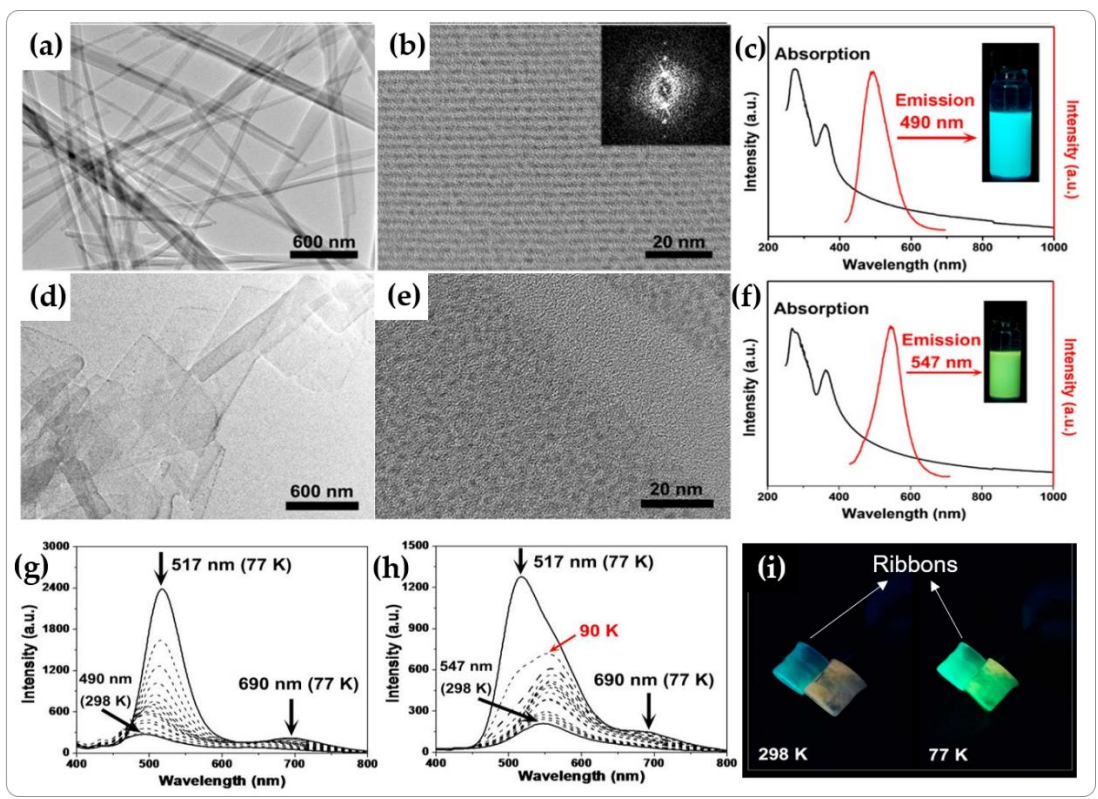

**Fig. 3.** TEM characterization and fluorescence spectra of CuNCs with different self-assembly morphology: (a-c) ribbons; (d-f) sheets. Variation of emission spectra of ribbons (g) and sheets (h) with 365 nm excitation from 77 to 298 K. Fluorescent images of the powder (i) of CuNCs self-assembly ribbons and sheets with 365 nm excitation at 298 and 77 K (reprinted from (Wu et al. 2015) with permission from American Chemical Society).

As an important new type of porous material, metal-organic frameworks (MOFs) are widely used in catalysis, energy storage, and separation (Maurin et al. 2017). Recently, a simple one-pot synthesis of GSH-CuNC and MOF complexes (CuNCs/MOFs) was reported for the first time (Fig. 4) (Han et al. 2018b). GSH-CuNCs were evenly distributed across the whole MOFs structure. The fluorescence intensity of CuNCs/MOFs increased about 35-fold more than that of individual GSH-CuNCs because of the MOFs skeleton that exhibited a space limiting effect. Additionally, the stability timeframe was greatly enhanced from 3 days to 3 months. The difference was that the fluorescence emission of the synthesized CuNCs/MOFs showed a blue shift of about 20 nm which might have been caused by the predominance of inter-Cu(I)···Cu(I) interactions over intra-interactions. However, when the pH of the solution changed, the synthesized complex exhibited both significant morphological differences and a drastic change in its fluorescence emission. At a lower pH value, the crystal structure of MOFs was easily destroyed. When the pH was lowered to about 3.5, the solution changed from turbid to clear, and the orange fluorescence emission of the solution disappeared with the blue emission occurring at 440 nm. When NaOH was added again, white turbidity appeared again with a recovery of the orange fluorescence emission. For this reason, the CuNCs/MOFs complex shows great potential as a pH probe. Notably, pH sensitivity also limits broader applications of MOFs-based CuNCs.

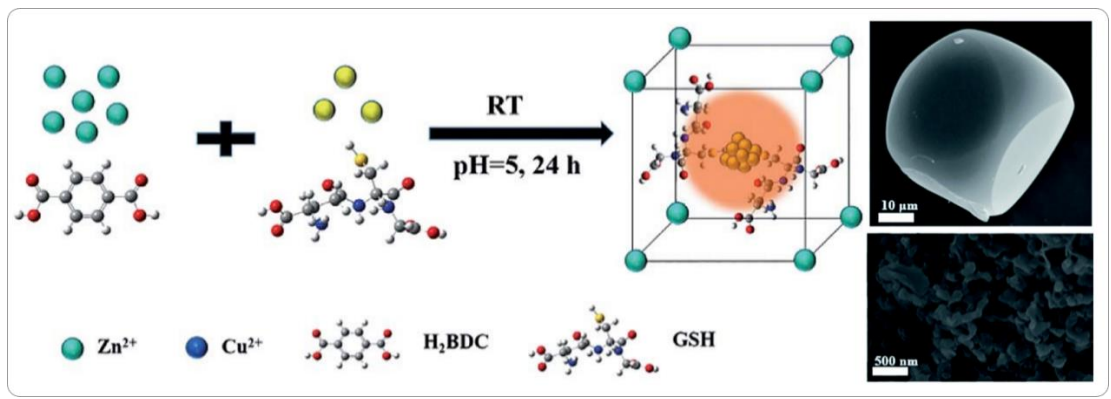

**Fig. 4.** Schematic illustration of the one-pot synthesis of CuNCs/MOFs composites. Insets show the SEM images of MOFs (upper) and CuNCs@GSH/MOFs composites (lower) (reprinted from (Han et al. 2018b) with permission from Royal Society of Chemistry).

*2.3. Solvent effect and AIEE*

2.3.1. Solvent effect

The ground and excited states of fluorescent molecules have different electron distributions, resulting in different dipole moments and polarizability (Liptay 1974). Different interactions among the ground state, the excited state, and the solvent molecule have a great influence on the fluorescence spectrum and intensity. In this regard, the same fluorescent material may have various fluorescence spectra in different solvents. For example, the red shift of fluorescence spectrum resulting from the solvent effect has been demonstrated by theoretical and experimental studies (Han et al. 2018c; Prakash et al. 2019).

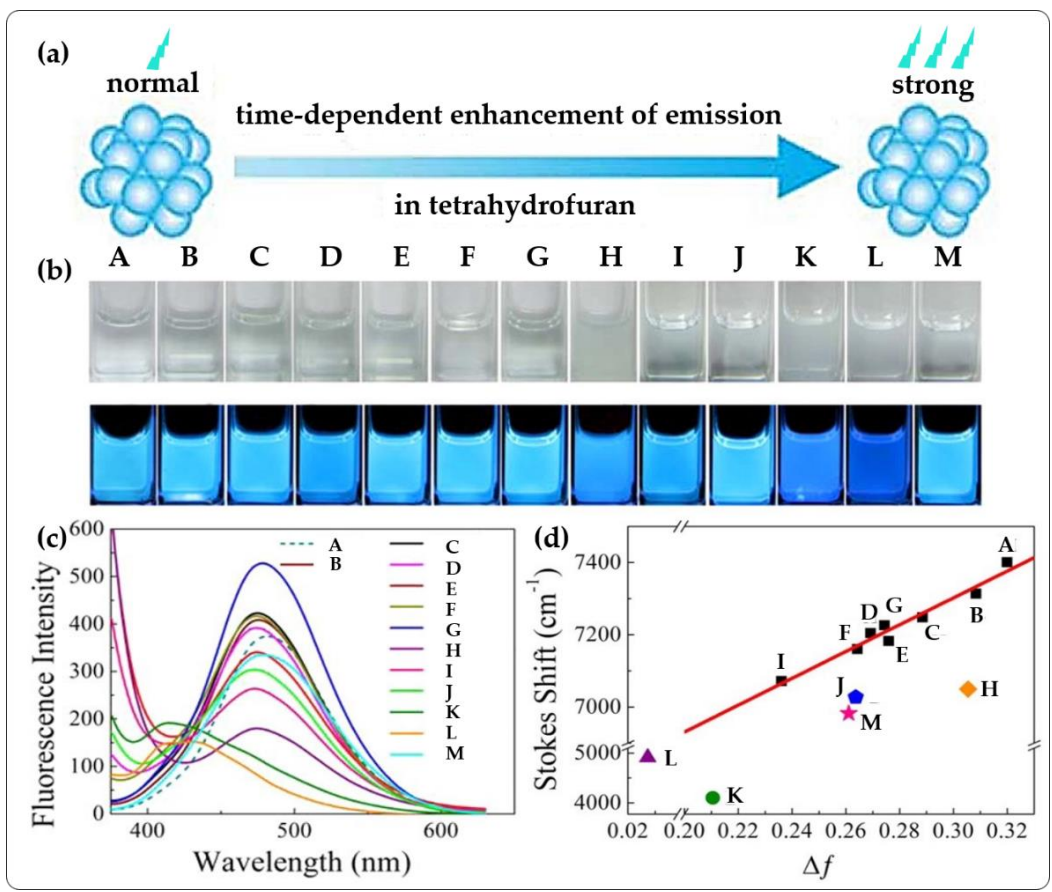

**Fig. 5.** (a) Fluorescence spectra of the PEI-capped Cu nanoclusters dispersed in different solvents; (b) Lippert plot of PEI-Cu nanoclusters dispersed in different solvents: (A) water, (B) methanol, (C) ethanol, (D) n-propanol, (E) isopropanol, (F) 1-butanol, (G) ethylene glycol, (H) acetonitrile, (I) N,N-Dimethylformamide, (J) DMSO, (K) THF, (L) 1,4methanol, and (M) EGME (reprinted from (Ling et al. 2015) with permission from American Chemical Society).

Ling et al. studied the solvent effect in 12 polar organic solvents, using polyethyleneimine (PEI)-terminated CuNCs as a research target (Ling et al. 2015). They found that the fluorescence spectra and characteristic absorption spectra of PEI-CuNCs in alcohol solvents were similar to those in water. However, a special phenomenon was observed in solvents such as acetonitrile, tetrahydrofuran (THF), dimethyl sulfoxide (DMSO), ethylene glycol monomethyl ether (EGME), and 1,4-dioxane. Namely, blue shifts in emission peaks were found in THF and 1,4-dioxane (Fig. 5a), and there were significant changes in their absorption spectra. The solvent effect was generally described according to the Lippert equation. The specific calculation method is shown below.

$$v_a - v_f = \frac{2(\mu^*-\mu)^2}{hca^3}\left(\frac{\varepsilon-1}{2\varepsilon+1} - \frac{n^2-1}{2n^2+1}\right), \qquad (1)$$

The different behavior on the Stokes shift was induced by the dielectric constant (ε) and refractive index (n). The $v_a$ and $v_f$ are the frequencies of the absorption and emission maxima, respectively. The μ* and μ are the dipole moments in excited-state and ground-state, respectively. c is the speed of light, h is Planck's constant, and *a* is the solute cavity radius. The orientation polarizability (Δf) is in parentheses, $v_a$ minus $v_f$ is the Stokes shift. Generally, the Stokes shift decreases when the refractive index increases, whereas an increase in dielectric constant leads to a larger Stokes shift (Wong et al. 1992). The author also gave the relationship between Stokes shift and Δf. As shown in Fig. 5b-d, as the orientation polarizability increased, the Stokes shift also increases. In addition, THF and 1,4-dioxane were more likely to form hydrogen bonds with water. The hydrogen bond affected the absorption and emission spectra if a hydrogen bond interaction occurred in the ground state.

The development of luminescent materials is indispensable in the process of human life and civilization. Aggregation-induced emission enhancement (AIEE) is a phenomenon that was accidentally discovered by Tang's group in 2001 (Hong et al. 2011). Researchers found that the AIEE effect was also applicable to the synthesis of nanoclusters (Jia et al. 2013; Jia et al. 2014; Kang et al. 2016). The aggregated NCs can change the ligand/ligand, ligand/metal, and metal/metal interactions, which affects the excited state relaxation kinetics. In the non-aggregated state, ligand molecules return electrons from excited state to the ground state by internal rotation or vibration (Goswami et al. 2016; Hong et al. 2011). In the state of aggregation, intramolecular rotation and vibration are greatly restricted, as the energy in the excited state can only be released by non-radiative transitions, which is manifested by a strong increase in fluorescence intensity. Strategically, the AIEE effect is used to increase the radiant energy transfer rate by inhibiting the ligand-related non-radiative relaxation of the excited state (Luo et al. 2012). It is worth mentioning that this strategy gives us insight into designing high-efficiency luminescent CuNCs. It has been found that many CuNCs have unique AIEE properties (Li et al. 2016a; Liu et al. 2018b; Zhou et al. 2019).

Recently, Subarna et al. (Maity et al. 2019) gave a clear explanation of solvent effect and the AIEE effect. $Cu_{34-32}(SG)_{16-13}$ NCs were successfully synthesized using GSH as the ligand. They found that the addition of ethanol greatly affected the fluorescence emission of CuNCs. The extent of aggregation was controlled by the ratio of water to ethanol. In aqueous solution, $Cu_{34-32}(SG)_{16-13}$ NCs with surface electronegativity could form hydrated shells and exhibit good solubility and stability. Due to the difference in polarity, the addition of ethanol could destroy the hydrated shell and cause the surface charge to be neutralized (Maity et al. 2019). The particles lacked stability and became closer to each other due to higher inter and intra NCs Cu (I)···Cu (I) hydrophilic interactions. Large aggregates gradually formed and showed enhanced fluorescence intensity.

### 2.3.2. pH-dependent AIEE

Fluorescence properties vary with pH, usually accompanied by the formation of hydrogen bonds or protonation/deprotonation processes leading to significant changes in the excitation (emission) spectra. Superior optical reversibility is the basis for the further use of pH-responsive probes. Su et al. (Su and Liu 2017) performed an in-depth study of pH-responsive CuNCs, using L-cysteine (Cys) as both the reducing and capping agents. They found that CuNCs showed a good reversible pH-responsive aggregation (from 1.0 to 7.0). As shown in Fig. 6, CuNCs aggregated with each other and emitted intense red fluorescence at a lower pH of 3.0. However, as the pH gradually increased from 3.0 to 7.0, the CuNCs aggregate in the solution dissolved, accompanied by a significant decrease in fluorescence emission. The same phenomenon occurred when the pH was reduced from 3.0 to 1.0. The AIEE behavior could be attributed to the formation of hydrogen bonds between carboxyl and amino groups of cysteine under acidic conditions (Brinas et al. 2008; Wang et al. 2013). The changes of CuNCs from monodisperse CuNCs to aggregates mainly involved the interaction of ligands on the surface and the hydrophilic interaction of Cu(I)···Cu(I). Specifically, it could be divided into two main factors. First, the enhanced cuprophilic interactions facilitated the radiative pathway from the excited-state relaxation dynamics. Second, the restriction of intramolecular vibration and rotation reduced the nonradiative relaxation of the excited states (Su and Liu 2017). The pH-responsive properties of CuNCs made it possible for them to be used in tumor-specific imaging. The authors also showed that the pH responsiveness allowed CuNCs to self-assemble with BSA and glucose oxidase (GOx) and that the assembles also had similar pH responsiveness. Interestingly, the synthesized GOx-CuNCs emitting red fluorescence exhibited excellent responsiveness to glucose. In the presence of GOx-CuNCs, glucose could be oxidized by oxygen to form $H_2O_2$, which quenched the luminescence of CuNCs with a detection limit of 1.5 μM in the concentration range of 5-100 μM. This strategy provides an important guideline for the design of water-soluble luminescent biomolecule-stabilized NC hybrid nanostructures.

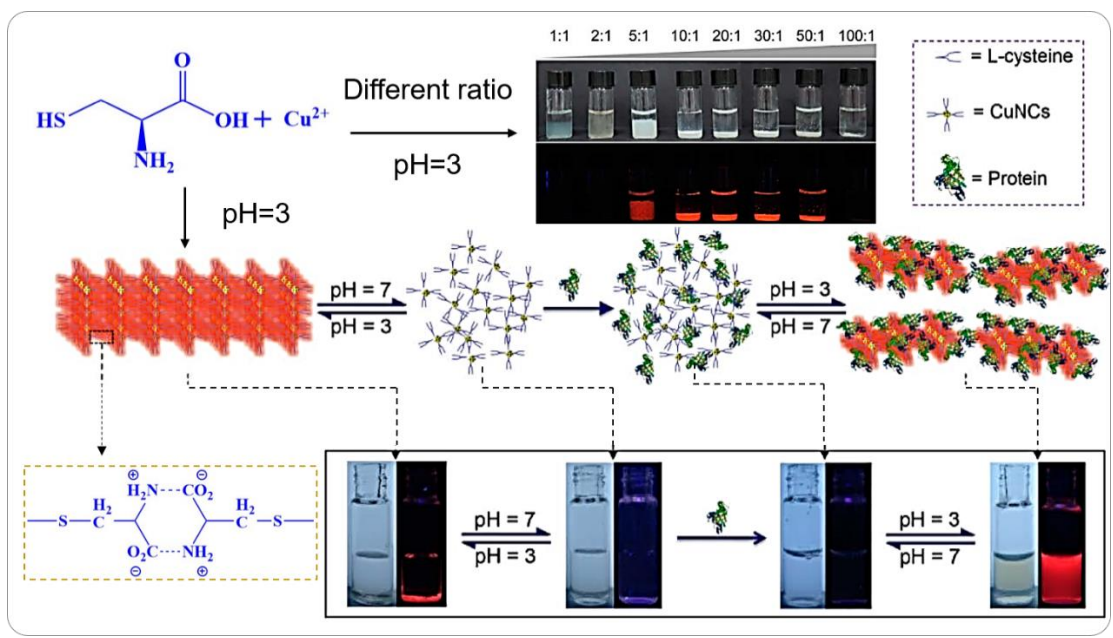

**Fig. 6.** Composition schematic of CuNCs and pH-induced AIEE phenomenon. Inset showed the photographs of the CuNC aggregates synthesized at different molar ratios of cysteine to $Cu^{2+}$ at room temperature, and the pictures were taken under room light (upper panel) and 365nm UV light (lower panel) (reprinted from (Su and Liu 2017) with permission from American Chemical Society).

2.3.3. Ion-induced AIEE

Cation-induced aggregation is one of main methods used for enhancing the emission efficiency of CuNCs. Generally, cation-induced aggregation can be attributed to the neutralization of surface charges on CuNCs or the formation of new bonds due to the high affinity.

Hu et al. identified that the fluorescence emission of dithiothreitol (DTT)-capped CuNCs could be significantly enhanced when $Al^{3+}$ was added into the solution (Hu et al. 2017). It was found that, with the increase in $Al^{3+}$ concentrations, CuNCs aggregated, and the resulting hydrodynamic diameter increased over 14-fold. Fundamentally, this phenomenon was observed for the following two reasons. First, the negative charge on the surface of CuNCs was neutralized by $Al^{3+}$ due to charge interaction, resulting in a decrease in zeta potential and the aggregation of CuNCs. Second, $Al^{3+}$ could combine with ligands on the surface of CuNCs to produce Al-O bonds, further promoting the aggregation of CuNCs. These two interactions gave rise to remarkable enhancement of fluorescence emission. Notably, this method was successfully applied to the detection of $Al^{3+}$ in food with a detection limit as low as 0.01 μM. Based on a similar mechanism, strong fluorescence enhancement was observed when $Zn^{2+}$ was added to GSH-CuNCs. In the presence of $Zn^{2+}$, the fluorescence intensity of CuNCs tripled, and the absolute quantum yield increased from 1.3% to about 6.2% (Lin et al. 2017). A good linear detection range was observed from 4.68-2240 μM and the limit of detection was about 1.17 μM. Due to the biocompatibility, CuNCs showed great advantages in $Zn^{2+}$ imaging of living cells. Moreover, Li et al. (Li et al. 2018) reported a method for detecting $Hg^{2+}$ based on induced fluorescence enhancement. They found that $Hg^{2+}$ could alter the structures of DNA. As shown in Fig. 7a, $Hg^{2+}$ induced single-stranded poly thymine (T) DNA to form a network structure via T−Hg$^{2+}$−T base pairs, resulting in the formation of reticulated DNA. Compared with poly(T) DNA, poly(T) DNA treated by Hg$^{2+}$ was more rigid. The reticular structure limited its non-radiative transitions, thus resulting in the formation of CuNCs with enhanced fluorescence. In this case, Hg$^{2+}$ could be easily detected by fluorescence enhancement. The method used exonuclease I (exo I) to digest excess poly(T) DNA. Due to the high selectivity and specificity of the enzyme, exo I can remove nucleotides from single-stranded DNA rather than reticular DNA. It could not only reduce background noise, but could also protect the enhanced signal. It is worth mentioning that this method has good selectivity and can be used for water quality testing. In addition, since the introduction of Hg$^{2+}$ and GSH is associated with a decrease in fluorescence intensity, the biosensor was in principle expected to be used to detect GSH.

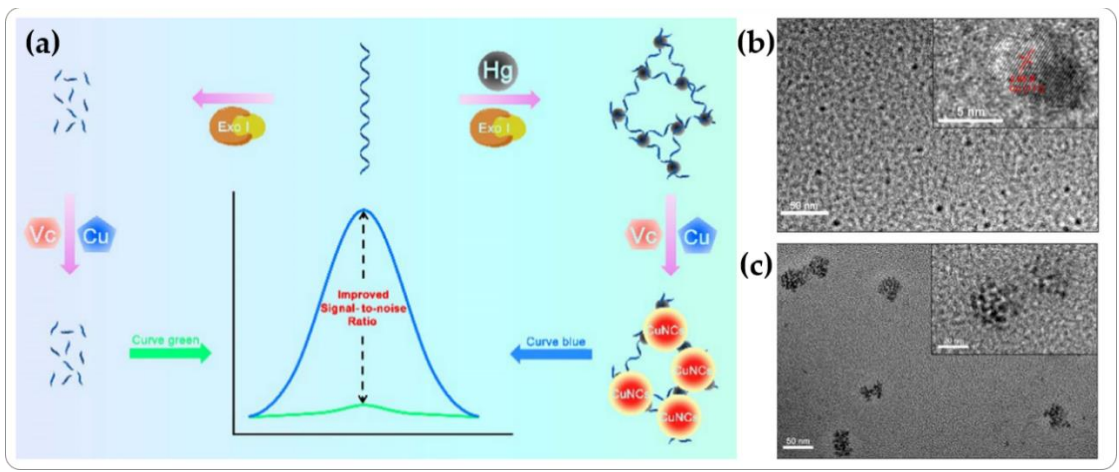

**Fig. 7.** Schematic illustration of Hg$^{2+}$ quantification based on fluorescence regulation of CuNCs via DNA template manipulation. (b) TEM image of CuNCs using poly(T) as the template. Inset: HRTEM image of CuNCs with a lattice spacing of 2.05 Å. (c) TEM image of CuNCs using Hg$^{2+}$-treated poly(T) as the template. Inset: Enlargement of one aggregate of CuNCs (reprinted from (Li et al. 2018) with permission from American Chemical Society).

Usually, aggregation-induced products are more likely to be large productions or network structures. Recently, it was reported that S$^{2-}$ was capable of inducing silk fibroin (SF) protected CuNCs (SF@CuNCs) to form uniform rod-shaped fluorescence-enhanced nanoparticles (Zhang et al. 2019a). The generation of aggregation was attributed to the interaction of S$^{2-}$ with the core of SF@CuNCs. Based on this finding, it could be extended to the detection of S$^{2-}$. The limit of the given detection was 0.286 μM in the linear concentration of 5-110 μM.

## 3. CuNCs for sensing

Selectivity, stability, detection range, and detection limit are the four important criteria when considering the feasibility of CuNCs as probes. Over the past few years, CuNCs have been used as intriguing ion probes because of their superior optical properties. By capping them with different ligands, CuNCs usually display powerful recognition capabilities, from small ions to large biological molecules. Strategically, the majority of detection methods follow a similar principle. The concentration of the target substrate can be determined based on the difference in fluorescence intensity before and after the detecting treatments.

*3.1. Ion detection*

3.1.1. Al$^{3+}$ detection

So far, it has been well documented that $Al^{3+}$ can enhance the fluorescence intensity of CuNCs. Boonmee et al. found that cysteamine capped CuNCs showed good responsiveness to $Al^{3+}$ (Boonmee et al. 2018). As the cysteamine has only two carbon atoms and the $-NH_2$ moiety is very close to the core cluster, a strong fluorescent emission might have happened once the target interactions occurred at the $-NH_2$ moiety. In the presence of $Al^{3+}$ with a small ion radius and high charge, CuNCs were induced to aggregate according to the coordination interactions. In the acidic solution, the presence of $Al^{3+}$ could cause a significant increase in the fluorescence intensity of cysteamine-CuNCs at 380 nm, while other studied metal ions did not produce this effect. With the increase of $Al^{3+}$ concentrations, the fluorescence intensity at 380 nm rose remarkably. The selectivity of CuNCs allowed it to be applied to determine $Al^{3+}$ concentration in drinking water samples, providing a working range of 1-7 μM with a low detection limit of 26.7 nM.

Differing from the above fluorescence enhancement principle, Pang et al. designed a selective fluorescent switching probe for the detection of $F^-$ using $Al^{3+}$ as a bridge (Pang et al. 2019). The red fluorescent CuNCs were synthesized with polythymidine as a template. In the presence of $Al^{3+}$, the fluorescence of CuNCs was greatly quenched due to the coordination interactions between $Al^{3+}$ and DNA. However, the addition of $F^-$ caused the desorption of $Al^{3+}$ from DNA and the formation of the $Al(OH)_3F^-$ complex, resulting in the restoration of fluorescence intensity. The proposed method was successfully applied to determine $F^-$ concentration in toothpaste, which delivered a working range of 2 to 150 μM and a low detection limit of 1 μM. Moreover, a similar method was used for the detection of creatinine (CRN) with GSH-CuNCs (Jalili and Khataee 2018). The detection limit is 0.63 μg/L in the range of 2.5-34 μg/L.

### 3.1.2. $Hg^{2+}$ detection

As a harmful heavy metal element, mercury exists in different forms in nature. Excessive exposure to an environment rich in Hg causes serious adverse effects to human body. Recently, Bhamore et al. synthesized CuNCs with bright blue fluorescence using curcuminoids as the template (Bhamore et al. 2018). They observed that, while the $Hg^{2+}$ was added to the CuNCs, the dispersed CuNCs aggregated to each other and caused fluorescence quenching due to the strong interaction between $Hg^{2+}$ ions and ligands on the surface of CuNCs. Based on this finding, a novel fluorescent $Hg^{2+}$ sensor was established, showing a linear range of 0.0005-25 μM with a detection limit of 0.12 nM at room temperature.

Liu et al. (Liu et al. 2019a) proposed a colorimetric method for the detection of $Mg^{2+}$ or $Pb^{2+}$ in solution. In the absence of $Pb^{2+}$ or $Hg^{2+}$, $H_2O_2$ could be decomposed to produce oxygen and water, which inhibited the oxidation of 3,3',5,5'-Tetramethylbenzidine (TMB) because of the catalase-like activity of metallothionein-stabilized CuNCs (MT-CuNCs). However, the catalase-like activity of the MT-CuNCs was converted into peroxidase-like activity in the presence of $Pb^{2+}$ or $Hg^{2+}$. As a result, $H_2O_2$ was catalytically decomposed to produce hydroxyl radicals that would react with TMB. TMB was oxidized to 3,3′,5,5′-tetramethylbenzidine diimine (oxTMB) and accompanied by changes in the color of the solution (Fig. 8). Taking advantage of this property, $Mg^{2+}$ or $Pb^{2+}$ could be easily detected in the UV spectrum. Under optimal conditions, a linear relationship was observed while detecting $Pb^{2+}$ in the concentration range of 707 nM to 96 μM with a detection limit of 142 nM. As for $Hg^{2+}$, a linear relationship was observed in the range of 97 nM to 2.325 μM and 3.10 to 15.59 μM with a detection limit of 43.8 nM.

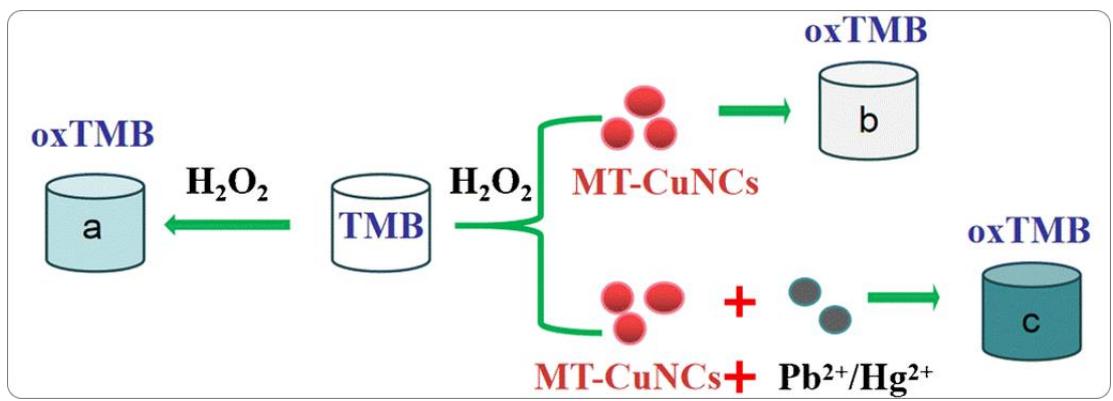

**Fig. 8.** Schematic diagram of MT-CuNCs for the Detection of $Pb^{2+}$ or $Hg^{2+}$. In the presence of $Pb^{2+}$ or $Hg^{2+}$, catalase-like activity of the MT-CuNCs was converted into peroxidase-like activity. $H_2O_2$ was catalytically decomposed to produce hydroxyl radicals that reacted with TMB (reprinted from (Liu et al. 2019a) with permission from Springer).

### 3.1.3. $Cu^{2+}$ detection

Copper participates in many physiological processes in living organisms involving the electron transfer catalysis of two oxidation states (Gawande et al. 2016). However, excessive exposure to high concentrations of $Cu^{2+}$ causes damage to the gastrointestinal tract, liver, and kidneys. Recently, Zhong et al. (Zhong et al. 2014) observed that the addition of $Cu^{2+}$ led to increased interactions between $Cu^{2+}$ and BSA, which quenched the fluorescence of BSA-CuNCs. By adding ethylene diamine tetraacetic acid to the system, $Cu^{2+}$ was removed from the surface of BSA-CuNCs, resulting in the restoration of fluorescence. Due to the lack of selectivity, the presence of $Hg^{2+}$ or $Fe^{3+}$ also led to the similar influences which limited its practicability. Based on the AIEE effect described above, Shen et al. (Shen et al. 2019) proposed a simple method for detecting $Cu^{2+}$ in solution. The sample to be tested was directly added to the aqueous solution of tetrahydrofuran containing GSH. $Cu^{2+}$ was reduced to form GSH-CuNCs, and aggregated into large particle aggregates with red emission. It is worth noting that the reaction time is 2 minutes with a detection limit of 0.17 μM.

### 3.1.4. $S^{2-}$ detection

Recently, a dual emission nanocomposite of CuNCs/CDs (carbon dots) was reportedly used as a ratiometric fluorescent probe for the detection of sulfide and gaseous $H_2S$ (Fig. 9a) (Wen et al. 2019). This nanocomposite presented two emission peaks at 469 nm and 622 nm, which were attributed to CDs and CuNCs, respectively. The addition of sulfides quenched the fluorescence of CuNCs, while the fluorescence intensity of CDs remained constant (Fig. 9b). After $S^{2-}$ treatment, CuNCs were converted into CuS, while CDs displayed a negligible change. As a result, the fluorescence of the probe changed from red to blue as the concentration of sulfide increased. As shown in Fig. 9c, a strong linear relationship could be observed in the concentration range of 2 to 10 ppb (26-128 nM) with a very low detection limit of 4.3 nM. In light of the unique color change of ratiometric method, portable fluorescent test paper was prepared by combining the nanocomposite with agar gel. Furthermore, the test paper had a good response to gaseous $H_2S$, which was of great importance for $H_2S$ detection in the environment (Fig. 9d). A similar structure has also been reported in other noble metals. Since dual-emitting probes are more resistant to complex sample environments (Wang et al. 2016a), they have great potential in biological applications.

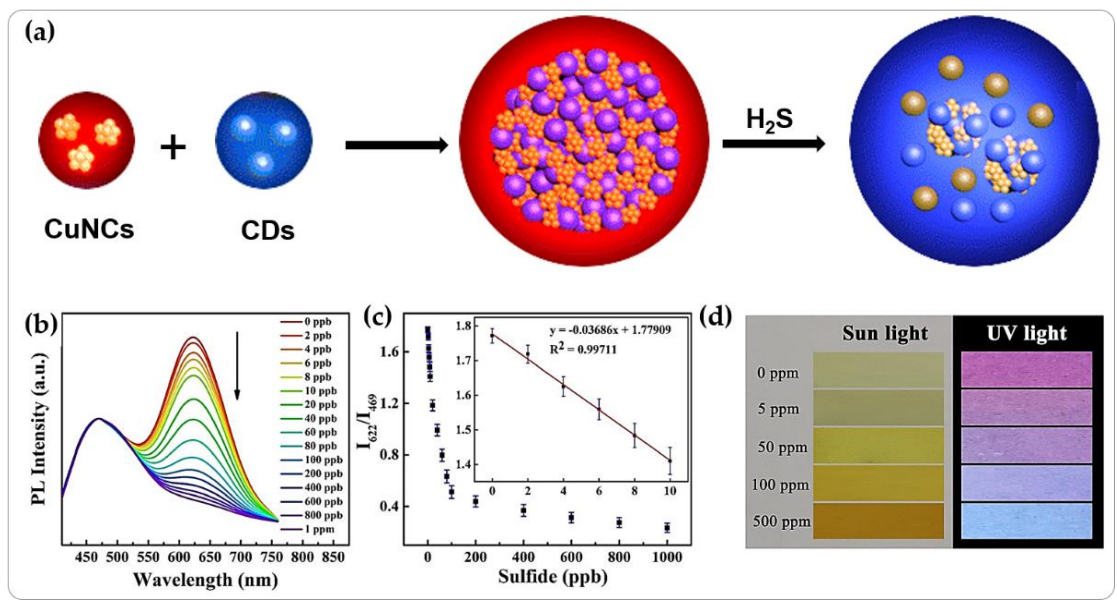

**Fig. 9.** (a) Schematic presentation of the synthesis of CuNCs-CDs dual-emission nano-assembly. (b) PL spectra of CuNCs-CDs dual-emission nano-assembly in the concentration of sulfide from 0 to 1 ppm; (c) The relationship between the ratio of fluorescence intensity ($I_{622}/I_{469}$) and sulfide. Inset: linear relationship of the ratio of fluorescence intensity ($I_{622}/I_{469}$) and sulfide from 2 to 10 ppb. (d) Photographs of CuNCs-CDs-agar fluorescent film in different concentration of gaseous $H_2S$ under sun light and UV light (reprinted from (Wen et al. 2019) with permission from Springer).

*3.2. Aromatic compound detection*

3.2.1. Trinitrotoluene (TNT) detection

Aparna et al. (Aparna et al. 2018) proposed that PEI-CuNCs could be used for detecting environmental TNT. The amino group of the PEI on the surface of CuNCs could chemically recognize TNT molecules by forming TNT derivatives and reducing the fluorescence intensity. It could be attributed to the following two reasons. First, the overlap of the excitation spectrum of CuNCs with the absorption spectrum of TNT led to the internal filter effect. Second, the emission spectrum of CuNCs overlapped with the absorption spectrum of the TNT derivative, resulting in a Forster resonance energy transfer (FRET) phenomenon. The formation of such a complex occurred with a significant color variation that made it possible to be used as a colorimetric sensor. Under optimal conditions, the detection limits of the probe through the fluorescent and colorimetric methods were 14 pM and 0.05 nM, respectively. In addition, PEI-CuNCs were successfully used in test paper to detect TNT vapor with a detection limit of 10 nM, which held great potential for detecting environment TNT.

3.2.2. Picric acid (PA) detection

Ravi et al. (Patel et al. 2018) found that pyridoxamine (PM) could specifically interact with GSH-CuNCs and thus adsorb to their surfaces, resulting in the formation of a new band near 410 nm in the fluorescence spectra. However, the addition of picric acid quenched the fluorescence both at 410 nm and 625 nm with high efficiency and selectivity (Fig. 10a) while other common metal ions or anions did not interfere with this detection. The possible mechanism for this might be the formation of hydrogen bonds between PA and PM on the surface of CuNCs. The fluorescence quenching at 410 nm was attributed to the resonance energy transfer from the excited PM-CuNCs to PA because of the overlap of the emission of PM-CuNCs and the absorbance of PA. The photo-induced electron transfer from PA to the excited PM-CuNCs (Malik et al.

2015) might explain the quenching of the red emission (Fig. 10b). The linear range given was 9.9-43 μM with a detection limit of 2.74 μM. In addition, highly selective PM-CuNC nano-assembly could be applied in real samples and was successfully used in the test paper. By modifying PM-CuNCs on cellulose strip test paper, the test paper showed distinctly different fluorescent colors while immersed in different concentrations (0.1 μM-1 mM) of PA solution (Fig. 10c-d). Since PA is an electron-deficient compound, it could easily interact with electron-rich groups. Other CuNCs have also been reported to present similar functions based on the fluorescence quenching mechanism (Shanmugaraj and John 2018; Zhang et al. 2018c). Cysteamine-coated CuNCs were prepared by Bao et al. (Bao et al. 2018) to show a superior detection range (1-80 μM) and detection limit (139 nM).

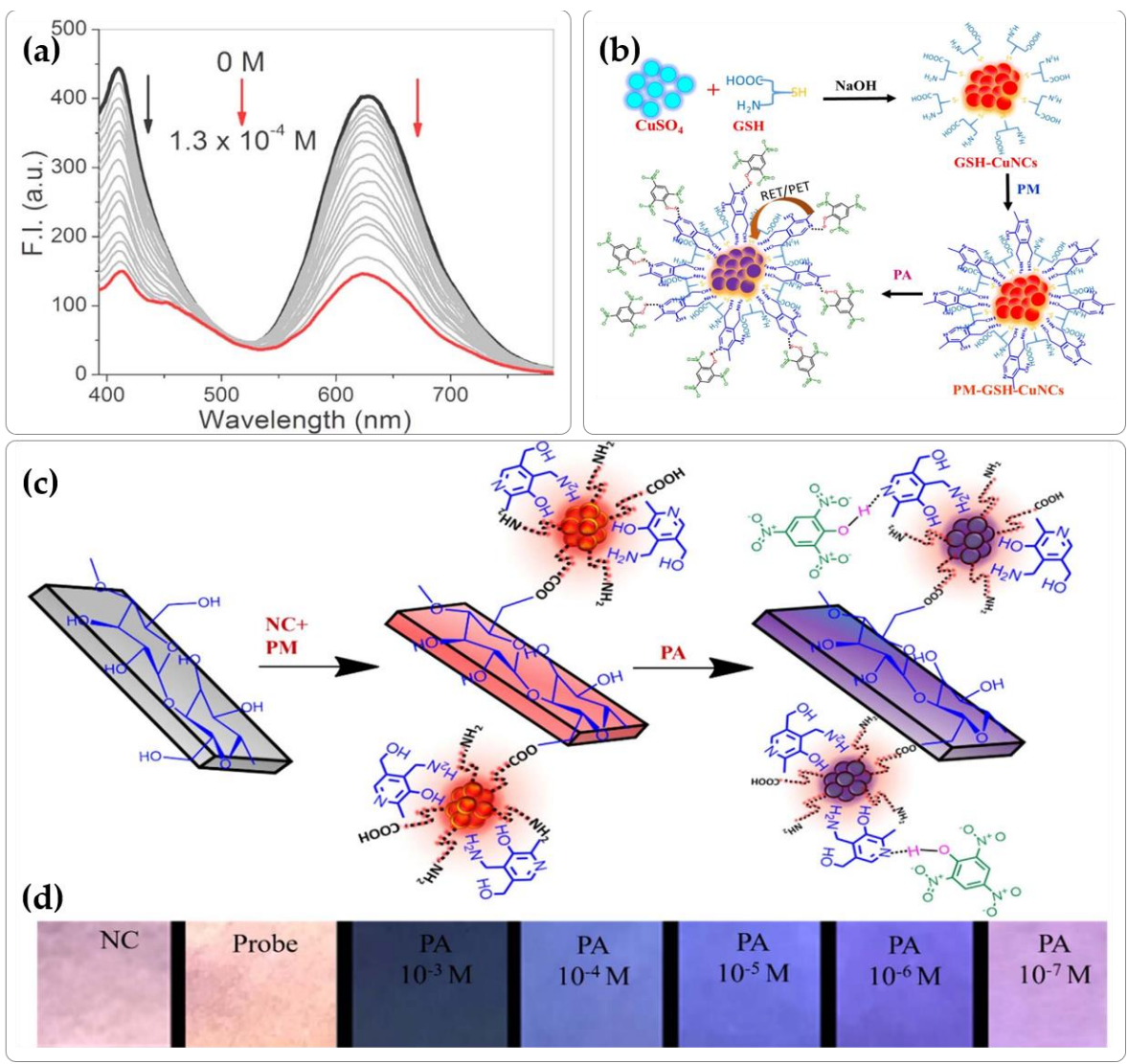

**Fig. 10.** (a) Fluorescence spectral changes of PM-CuNCs upon addition of increasing PA. (b) Schematic representation for the synthesis of GSH-CuNCs and its interactions with PM and PA. (c) Schematic diagram of CuNCs used in paper strips. (d) Fluorescence color changes of the modified paper strips upon interaction with different concentrations of PA (0.1 μM - 1 mM) observed under UV light at 365 nm (reprinted from (Patel et al. 2018) with permission from Elsevier).

3.2.3. M-dinitrobenzene and quinoline yellow detection

Based on AIEE mechanism, fluorescent Cys-CuNCs were used for the detection of m-dinitrobenzene (m-DNB) (Hambarde et al. 2019). Among different nitro-aromatics, m-DNB significantly enhanced the fluorescence of Cys-CuNCs in PBS buffer (pH 9.0). The detection limit was 0.13 μM in the range of 1.3-990 μM. Additionally, Cys-CuNCs were reported to be used for the detection of quinoline yellow, a food coloring agent (Sivasankaran et al. 2019). The addition of quinoline yellow quenched the fluorescence accompanied by the formation of new compounds, which could be observed in the UV spectrum. The linear detection range was 0.2-5.5 μM with a detection limit of 0.11 μM.

### 3.3. Neurotransmitter molecular detection

### 3.3.1. Dopamine detection

Dopamine acts as a neurotransmitter to regulate a variety of physiological functions of the central nervous system (Basu and Dasgupta 2000). For example, dopamine system dysregulation involving Parkinson's disease, schizophrenia, Tourette syndrome, etc. Miao et al. (Miao et al. 2018) found that BSA-CuNCs could selectively recognize dopamine. The fluorescence intensity of BSA-CuNCs rose with the increase of dopamine concentration at pH 7.4. Under optimized conditions, the fluorescence intensity of BSA-CuNCs showed strong linearity from 0.5 to 50 μM with a detection limit of 0.28 μM. It is worth noting that Aparna and colleagues (Aparna et al. 2019b) recently discovered that CuNCs in the presence of $H_2O_2$ had a faster response to DA. Furthermore, they developed a test paper for direct detection of dopamine. The working range was 0.1-0.6 nM with a low detection limit of 0.024 nM.

He et al. (He et al. 2018) proposed a new method for the detection of dopamine. CDs-CuNCs with dual emission were successfully designed and prepared. The novel complex consisting of 3-aminophenylboronic acid (APBA) modified CDs, and red-emitting BSA-CuNCs were constructed following a carbodiimide-activated coupling strategy. Since dopamine could specifically couple with APBA, the addition of dopamine caused the electrons to be transferred from CDs to dopamine, resulting in the fluorescence quenching (centering at 440 nm) of CDs in the complex, while the fluorescence (at 640 nm) of CuNCs remained constant. One important advantage of this dual emission system was the obvious fluorescence color change during the test, suggesting that the approximate sample concentration could be evaluated by the naked eye. This method provided a working range of 0.1-100 μM with a detection limit of 32 nM. The fluorescent probe could efficiently detect dopamine in human serum samples.

### 3.3.2. Carbamazepine detection

Recently, cetyltrimethylammonium bromide (CTAB) coated CuNCs were designed as a probe for the determination of carbamazepine (CBZ). In the study, CTAB served as the linking reagent, which could interact with CBZ to enhance fluorescence by blocking the non-radiative $e^-/h^+$ recombination defect sites on the surface of CuNCs (Hatefi et al. 2019). Specifically, a new emission peak appeared in the fluorescence spectrum at 480 nm, while the intensity at 415 nm remained unchanged in the presence of CBZ. Based on the relationship between fluorescence intensity and concentration, the concentration of CBZ in the sample could be easily obtained. The method gave a good linear relationship in the concentration range of 0.2-20 μg/mL with a detection limit of 0.08 μg/mL.

### 3.4. Biological enzyme detection

### 3.4.1. Uracil-DNA glycosylase (UDG) detection

Ling et al. (Ling et al. 2019) proposed a new method to detect UDG, which might exist in some DNA viruses. As shown in Fig. 11a, a double-stranded DNA (dsDNA) consisting of uracil bases (UB) and trigger primer (TP) with uracil damage was designed as a substrate. UDG identified and removed uracil damage in the substrate to expose abasic (AP) sites, which worked with exonuclease III (Exo III) to cause hydrolysis of dsDNA containing poly-adenine (PA) and poly-thymine (PT), and release of the PT sequence. After that, the TP sequence was used as a template to synthesize red fluorescent CuNCs. However, 4',6-diamidino-2-phenylindole (DAPI) interacted with dsDNA (UB-TP sequences and PA-PT sequences) to form a complex with blue fluorescence in the absence of UDG. Based on the relationship between fluorescence difference and UDG concentration, this new method could be applied to detect UDG. As shown in Fig. 11b, the experimental phenomenon was consistent with the fluorescence spectra. The increase in UDG content caused an increase in fluorescence intensity at 602 nm and a decrease at 452 nm. Conversely, a decrease in UDG content led to a decrease in fluorescence intensity at 602 nm and an increase at 452 nm. Within a certain range, the ratio fluorescence (RF) exhibited a good linear relationship with the concentration of UDG (Fig. 11c). This method gave a detection limit of 0.05 U/L, which might provide a potential platform for clinical diagnosis.

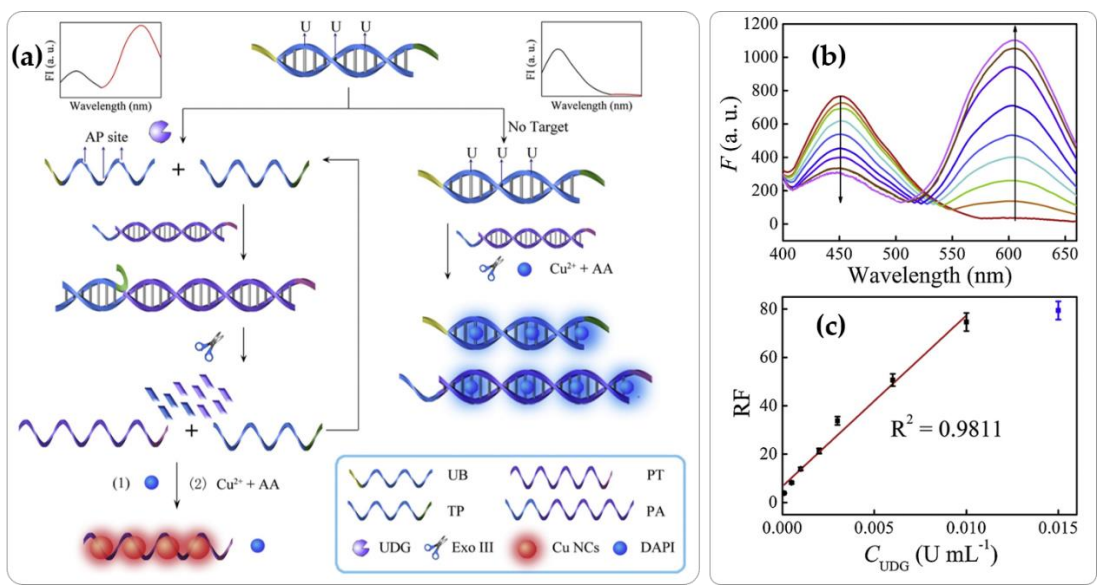

**Fig. 11.** (a) Schematic Diagram of CuNCs for the Detection of Uracil-DNA Glycosylase (b) Emission spectra of the system in the presence of various concentrations of UDG. (c) A linear relationship of RF, (F602/F°602)/(F452/F°452), versus the concentration of UDG. The concentrations of UDG are $1.0 \times 10^{-4}$, $5.0 \times 10^{-4}$, $1.0 \times 10^{-3}$, $2.0 \times 10^{-3}$, $3.0 \times 10^{-3}$, $6.0 \times 10^{-3}$, 0.01, 0.015 U/mL, respectively (reprinted from (Ling et al. 2019) with permission from Elsevier).

3.4.2. Pyrophosphatase detection

Liu and colleagues (Liu et al. 2018a) discovered that tannic acid (TA) coated CuNCs (TA-CuNCs) could be used to detect the activity of pyrophosphatase (PPA). Interestingly, it was found that $Fe^{3+}$ quenched the fluorescence of TA-CuNCs due to the interaction between TA and $Fe^{3+}$. Because of the stronger affinity between pyrophosphate (PPi) and $Fe^{3+}$, the addition of PPi removed $Fe^{3+}$ from the surface of CuNCs, resulting in the restoration of fluorescence. Furthermore, the PPi was decomposed into two phosphate ions when PPA was added to the solution, causing the fluorescence of the TA-CuNCs to be quenched again. Taking advantage of the "on-off-on" strategy, TA-CuNCs were successfully used for the detection of PPA in human serum samples with a detection limit of 0.19 U/L in the range of 0.5-18 U/L.

In addition, it was also reported that GSH-CuNCs exhibited a similar performance, which was used for detecting pyrophosphatase and alkaline phosphatase (Geng et al. 2019; Ye et al. 2019). The difference was that $Fe^{3+}$ was replaced by $Al^{3+}$, which could enhance the fluorescence of GSH-CuNCs mainly due to the

AIEE effect. Consequently, the detection limit and detection range were also different due to the replacement of the ligand. The linear ranges for the detection of pyrophosphatase and alkaline phosphatase were 3-40 U/L and 0.5-25 U/L, respectively; the detection limits were 1.3 U/L and 0.15 U/L, respectively.

### 3.4.3. Micrococcal nuclease detection

Staphylococcus aureus is one of the common etiological agents of food-borne illness. It often exists in blood vessels leading to different kinds of diseases (Suaifan et al. 2017). Micrococcal nuclease (MNase) as a non-specific endonuclease can digest DNA rich in adenine-thymidine or adenine-uracil base sequences. It was reported that the content of MNase could be used to directly assess the pathogenicity of staphylococcus aureus (Lagace-Wiens et al. 2007). As a consequence, it is of great importance to detect MNase simply and quickly. Since the dsDNA could be enzymatically digested into short oligonucleotide fragments or single nucleotides, the fluorescence of dsDNA-CuNCs was quenched in the presence of MNase. Guided by this sensing strategy, Qing et al. (Qing et al. 2019a) designed a dsDNA-CuNCs for the detection of MNase. The detection limit was 1 U/L in the range of 1-50 U/L.

### *3.5. Vitamin detection*

### 3.5.1. Folic acid (Vitamin $B_9$) detection

Ovalbumin (OVA)-stabilized CuNCs (OVA-CuNCs) with excellent water solubility and biocompatibility were reported by Li and his colleagues (Li et al. 2019a). It was found that the fluorescence of OVA-CuNCs was quickly quenched in the presence of folic acid (FA). Based on the static quenching mechanism, OVA-CuNCs were further used to detect FA with high selectivity. This method has been successfully applied to detect FA in fruits and vegetables and has worked well in the range of 0.5-200 μM with a detection limit of 0.18 μM. Earlier, Yang et al. (Yang et al. 2018) reported that OVA-CuNCs could be used to detect both Vitamin $B_1$ and doxycycline. In their study, the fluorescence of OVA-CuNCs was quenched by Vitamin $B_1$ and restored by the addition of doxycycline because of the different affinity with OVA-CuNCs. This method gave a linear range of 1-1000 nM for the detection of $VB_1$ and 1-1000 μM for detecting doxycycline. The detection limits were 380 pM and 270 nM, respectively.

Recently, a novel method for the determination of FA and nitrite was developed using Cys-CuNCs based on the reaction between diperiodate (III) (DPA) and FA (Han and Chen 2019). In this system, FA reacted with DPA and the product exhibited weak fluorescence. However, the fluorescence was significantly enhanced in the presence of Cys-CuNCs. The subsequent addition of nitrite quenched the fluorescence, as Cys-CuNCs would react with nitrite, resulting in the formation of a nonfluorescent ground-state complex. In addition, the reaction occurred with chemiluminescence, which was consistent with the fluorescent signal. The chemiluminescent signal was collected and was further amplified by the detector. This method was applied to detect nitrite in water and pickled vegetables. The detection limit was 0.0954 μM in the range of 1-80 μM.

### 3.5.2. Rutin (Vitamin P) detection

Flavonoids are known for their significant scavenging properties on oxygen radicals. As a natural flavonoid derivative, rutin is commonly found in many vegetables and fruits (Yang et al. 2008). Based on its good antioxidant properties, rutin has been used in cancer prevention and protection of nerves, and cardiovascular systems and cerebrovascular systems (Ganeshpurkar and Saluja 2017). Wang et al. (Wang et al. 2018a) demonstrated that rutin can effectively quench the fluorescence of BSA-CuNCs. This phenomenon was attributed to the formation of hydrogen bonds and the interactions between rutin and ligand. In this case, electrons were shifted from CuNCs to rutin, resulting in fluorescence quenching. Based on this finding, fluorescent BSA-CuNCs were applied for rutin detection. This method worked in a linear range of 0.1-100 μM, and the detection limit was 0.02 μM.

## 3.6. In situ *generation of DNA-CuNCs for detection*

DNA nanopreparation has attracted great interest as a template for bottom-up due to its excellent properties, including nanoscale geometry, programmable and artificial synthesis, DNA-metal ion interaction, and powerful molecular recognition capabilities (Qing et al. 2019b). Thymidine on the DNA template plays an important role in the formation of DNA-CuNCs. Fluorescent DNA-CuNCs with a special designed sequence have broadened prospects for staining and biosensing of ions, DNA, RNA, enzymes, proteins, small molecules, etc (Cao et al. 2019).

Recently, a dual-signal strategy based on small molecule-protein interactions was successfully used for detection of small molecules and protein solids. As shown in Fig. 12a, this method started from the preparation of $Fe_3O_4$@streptavidin complex by coating streptavidin on the surface of magnetic $Fe_3O_4$ nanoparticles. At the same time, biotin was linked to single-stranded DNA (ssDNA) by covalent bonds to form a biotin-ssDNA structure and then bound to a magnetic complex due to the specific interaction between biotin and streptavidin. In the presence of deoxy-ribonucleoside triphosphate (dTTP), the biotin-ssDNA was elongated to produce the long sequence of poly (T) with the aid of terminal deoxynucleotidyl transferase (TdT). Ideally, the magnetic nanoparticles bound to the elongated DNA precipitated under the force of magnetism. The precipitate treated with $Cu^{2+}$ and ascorbic acid emitted intense fluorescence, while the supernatant did not. However, it was quite different when the concentrations of either small molecules or proteins changed. Thus, the system can be easily utilized to detect streptavidin and biotin based on different competition mechanisms. It was observed that fluorescence intensity of the supernatant increased rapidly as the concentration of streptavidin increased, but the fluorescence intensity of precipitate almost maintained constant at 200 nM, since the elongated biotin-ssDNA achieved saturation in combination with streptavidin. Moreover, the fluorescence intensity of precipitate did not increase until the concentration of streptavidin increased to more than 200 nM. It was found that there was a good linear relationship between the ratio of fluorescence intensity of supernatant to the logarithm of the fluorescence intensity of precipitate and the streptavidin concentration in the range of 0-200 nM (Fig. 12b). In this way, we can easily obtain the concentration of streptavidin with a detection limit of 0.47 mM. A similar situation occurred when the detection target changed to the small molecule biotin. Biotin can competitively bind to magnetic $Fe_3O_4$@streptavidin, resulting in a strong fluorescence intensity in the supernatant. A linear range was obtained from 10-1000 nMwith a detection limit of 3.1 nM (Fig. 12c).

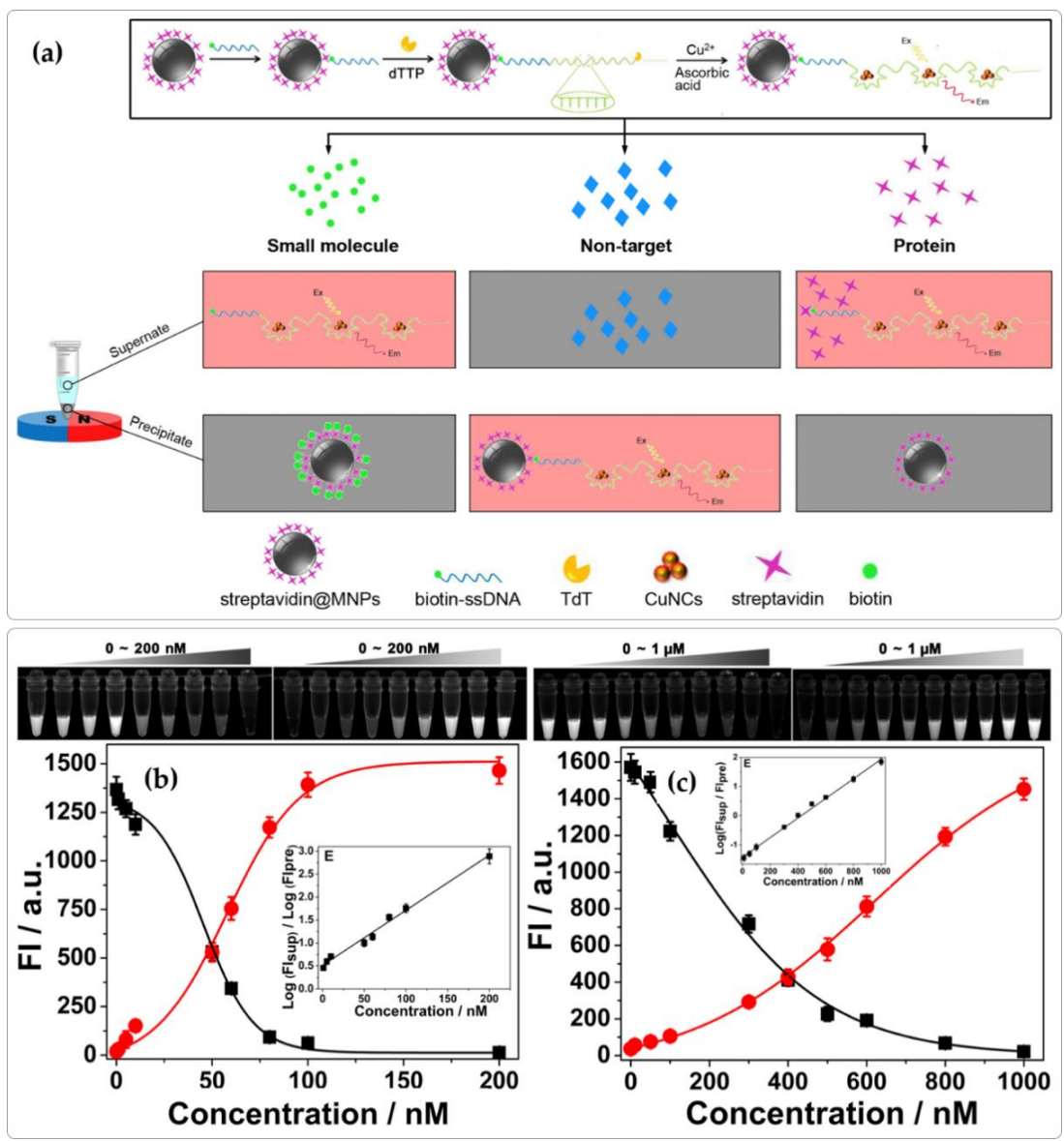

**Fig. 12.** (a) Schematic illustration of the principle for the detection of small molecules and proteins based on magnetic separation and highly fluorescent CuNCs. Fluorescence differences occurred at different locations in the tube based on different competition mechanisms. (b) Fluorescent intensity at 617 nm of the precipitate and the supernatant vs. the concentration of streptavidin. Inset shows the ratio of the logarithm of the fluorescent intensity of the supernatant to that of the precipitate vs. the concentration of streptavidin. Pictures above showed the fluorescence images of the precipitate (left) and the supernatant (right) with increased concentration of streptavidin. (c) Fluorescent intensity at 617 nm of the precipitate and the supernatant vs. the concentration of biotin. Inset shows the ratio of the logarithm of the fluorescent intensity of the supernatant to that of the precipitate vs. the concentration of biotin. Pictures above show the fluorescence images of the precipitate (left) and the supernatant (right) with increased concentration of biotin (reprinted from (Cao et al. 2017) with permission from Elsevier).

By *in situ* generation of CuNCs as an electro-chemiluminator, Liao et al. (Liao et al. 2018b) successfully constructed an ultrasensitive biosensor for microRNA-21 detection. As shown in Fig. 13a, the presence of

microRNA-21 initiated a cascade of reactions that produced a large number of secondary targets (T2) to open the hairpin DNA1 chain immobilized on the TiO$_2$@Pt interface. Then, a large number of AT-rich dsDNA was formed with DNA2 and DNA3 through the hybridization chain reaction, which combined with Cu$^{2+}$ *in situ* to generate electro-chemiluminescent CuNCs. In this study, TiO$_2$ was used to accelerate the reduction of S$_2$O$_8^{2-}$ to enhance the electrochemical luminescence (ECL) properties of CuNCs (Fig. 13b). The assay showed good selectivity and has been successfully used to detect microRNA-21 in human cervical cancer and human breast cancer cell lysates. A good linear relationship occurred in the range of 0.1 pM - 100 pM with a detection limit of 0.01905 pM (Fig. 13 c-d).

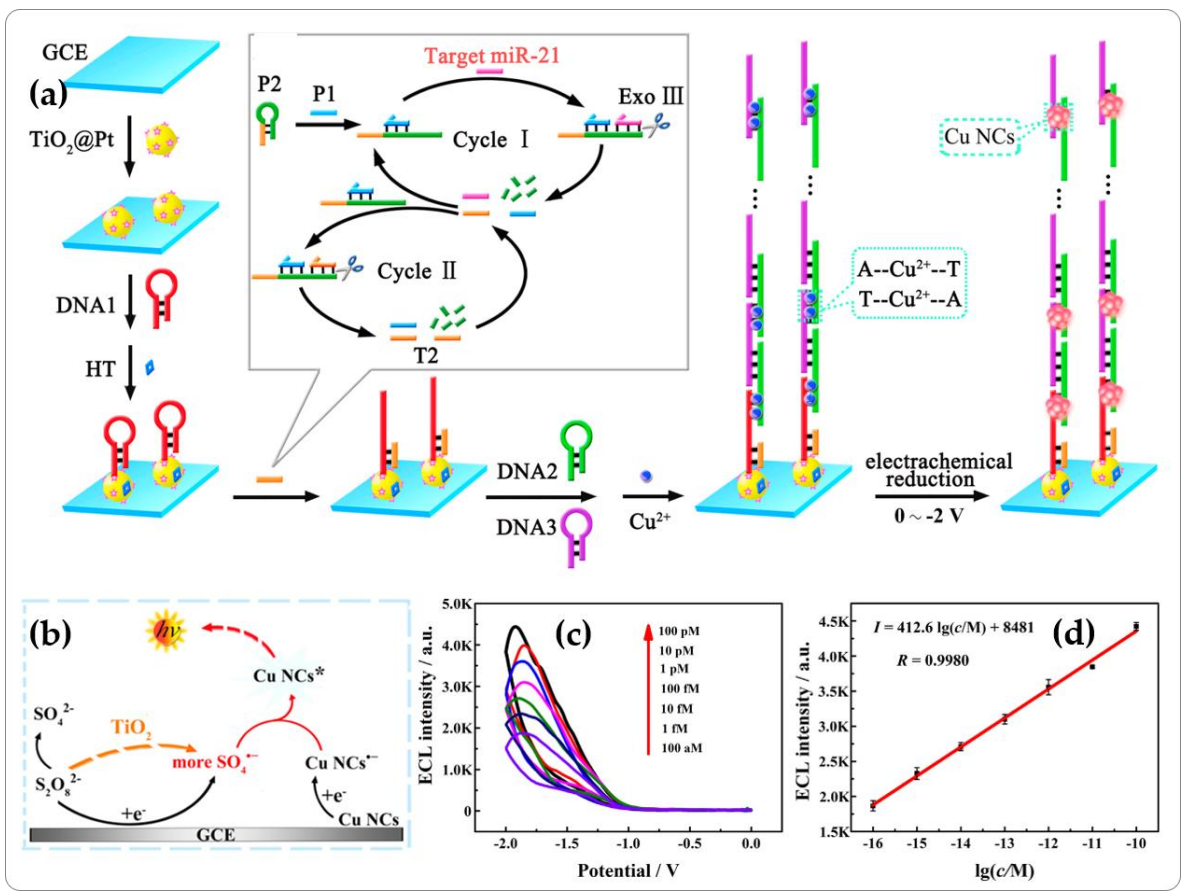

**Fig. 13.** (a) Schematic diagrams of the biosensor establishment process for the ultrasensitive determination of microRNA-21. Inset shows exo III-assisted cascade signal amplification process. (b) ECL mechanism of coreaction accelerator system with the promotion of TiO$_2$. (c) ECL intensity-voltage curves of the biosensor under different microRNA-21 concentrations. (d) Linearity curve for biosensors for microRNA-21 assays (reprinted from (Liao et al. 2018b) with permission from Elsevier).

### 3.7. Others detection

#### 3.7.1. Glucose detection

A new method has been developed to detect glucose based on the reaction between glucose and glucose oxidase (Gou et al. 2019). In the reaction system, sodium alginate, CaCO$_3$ nanoparticles (about 80 nm), and GSH-CuNCs were evenly distributed in the aqueous solution by ultrasonic treatment. When H$^+$ was added to the mixed system, CaCO$_3$ decomposed rapidly to produce Ca$^{2+}$. The generated Ca$^{2+}$ enhanced the fluorescence intensity of CuNCs in two ways. First, Ca$^{2+}$ crosslinked the alginate chain into a gel network to make the system more viscous; the vibration and rotation of molecules in the network were limited. Therefore, the fluorescence intensity was enhanced by inhibiting the non-radiative relaxation. Second, negative charge on the surface of GSH-CuNCs was neutralized by $Ca^{2+}$, resulting in aggregation. Taking advantage of this $H^+$-driven fluorescence enhancement, GOx was added into the system, and the mixed system was applied for the detection of glucose. Once glucose was present in the system, $H^+$ was quickly released by glucose and the GOx reaction, which triggered the above reaction (Fig. 14). Therefore, glucose was detected based on the relationship between fluorescence intensity and the concentration of glucose. In the study, a good linear relationship was attained in the range of 0.1-2 mM with a very low detection limit of 32 μM. However, the entire detection process took about 210 minutes, which may limit the application of CuNCs in Glucose detection.

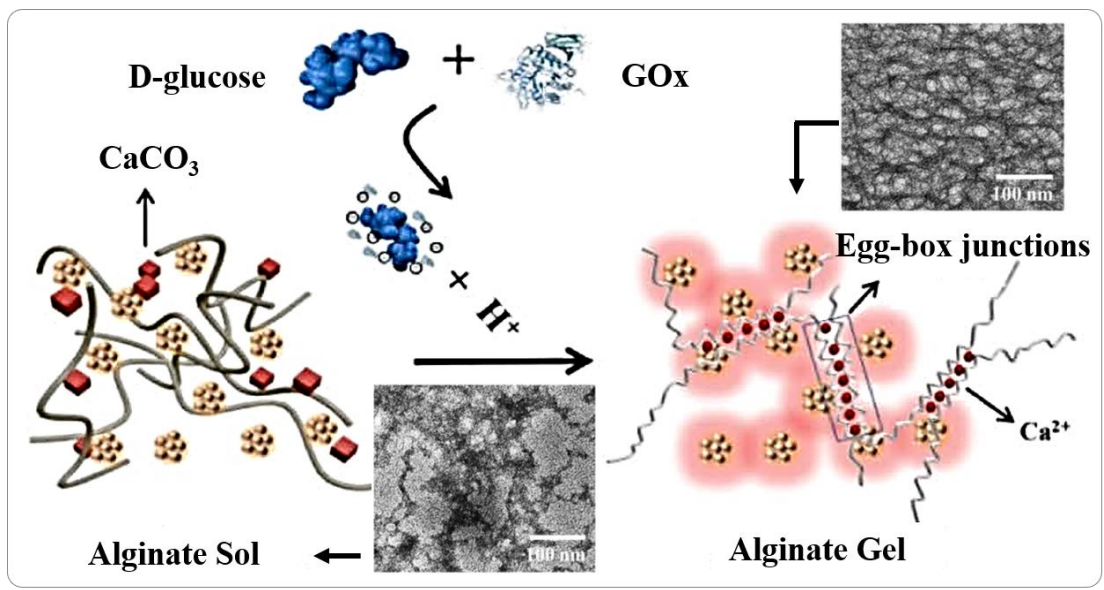

**Fig. 14.** Schematic illustration of the glucose sensor using hybrid CuNCs-alginate. Once glucose was present within the system, $H^+$ was quickly released to react with $CaCO_3$. $Ca^{2+}$ crosslinked the alginate chain into a gel network resulting in the fluorescence enhancement. Insets show the TEM images of pure alginate and the $Ca^{2+}$-triggered gel (reprinted from (Gou et al. 2019) with permission from American Chemical Society).

3.7.2. Histamine detection

The detection of histamine is imperative, because it plays an important role in nueroinflammation and microglial regulation (Wright et al. 2017). Histamine can cause an increase in blood vessel permeability due to multiple chronic and life-threatening inflammatory conditions, such as asthma and anaphylactic shock (Mikelis et al. 2015). It was reported that red fluorescent CuNCs coated with 2,3,5,6-tetrafluorothiophenol (TFTP) showed good responsiveness to histamines (Han et al. 2018a). The fluorescence intensity of CuNCs gradually decreased with the addition of histamine. The fluorescence quenching resulted from the strong interaction between the metal and the imidazole groups, resulting in electron transfer from the copper core to the ligands accompanied by the formation of electron-deficient sites on copper atoms. Under optimal conditions, the effective monitoring range of this method was 0.1 to 10 μM, with a detection limit of 60 nM. In addition, the prepared luminescent test strips containing TFTP-CuNCs were successfully used in the test of histamine content in fish and shrimp red wine.

3.7.3. Bilirubin detection

Bilirubin serves a vital role in antioxidants, and the lack of bilirubin may lead to coronary heart disease and anemia (Chou and Syu 2009). Recently, Ramar and his colleagues (Rajamanikandan and Ilanchelian 2019) reported red-emitting CuNCs, which could be effectively used for detection of bilirubin in clinical samples. CuNCs with excellent biocompatibility were stabilized by human serum albumin (HSA), which may avoid the potential immunologic reaction in human. In the study, the formation of non-fluorescent complexes of HSA-CuNCs and bilirubin led to a decrease in fluorescence intensity. This method exhibited a good linear relationship over two concentration ranges, 1.25-7.5 μM and 5-28.75 μM, and the limits of detection were 35 nM and 145 nM, respectively.

Over the past five years, CuNCs have achieved satisfactory detection for various analytes, mainly ions and virous biomolecules (proteins, nucleic acids, etc.). This review has briefly described some of the analysis methods as shown in Table 1 below.

**Table 1.** A summary of photophysical properties and analytical applications of CuNCs.

| Fluorescence sensing | Types of CuNCs | Maximum $\lambda_{em}$/nm ($\lambda_{ex}$)/nm | Quantum yield (QY)/% | Linear range | LOD | Ref. |
|---|---|---|---|---|---|---|
| $Al^{3+}$ | Cysteamine-CuNCs | 430 (330) | / | 1-7 μM | 26.7 nM | (Boonmee et al. 2018) |
| $Al^{3+}$ | DNA-CuNCs | 640 (340) | / | 2-150 μM | 1 μM | (Pang et al. 2019) |
| $Hg^{2+}$ | Curcuminoids-CuNCs | 440 (365) | 7.2% | 0.0005-25 μM | 0.12 nM | (Bhamore et al. 2018) |
| $Hg^{2+}$ | MT-CuNCs | / | / | 97 nM-2.325 μM and 3.1-15.59 μM | 43.8 nM | (Liu et al. 2019a) |
| $Hg^{2+}$ | Poly(T) DNA-CuNCs | 650 (340) | / | 50 pM-500 μM | 16 pM | (Li et al. 2018) |
| $Hg^{2+}$ | dsDNA-CuNCs | 595 (570) | / | 0.04-8 nM | 4 pM | (Zhang et al. 2018a) |
| $Cu^{2+}$ | GSH-CuNCs | 615 (330) | 1.2% | / | 0.17 μM | (Shen et al. 2019) |
| $Pb^{2+}$ | MT-CuNCs | / | / | 707 nM-96 μM | 142 nM | (Liu et al. 2019a) |
| $Ca^{2+}$ | CuNC@AF660 | 488 (590, 690) | 20% | 2-350 μM | 220±11 nM | (Liu et al. 2019b) |
| $Zn^{2+}$ | GSH-CuNCs | 340 (~605) | 1.3% | 4.68-2240 μM | 1.17 μM | (Lin et al. 2017) |
| $Fe^{3+}$ | Albumin chicken egg-CuNCs | 340 (280) | / | 0.2-100 μM | 0.0234 μM | (Huang et al. 2018) |
| $S^{2-}$ | GSH-CuNCs | 607 (360) | / | 0.5-20 μM | 0.5 μM | (Liao et al. 2018a) |

| Analyte | CuNCs | λem (λex) nm | QY | Linear range | LOD | Reference |
|---|---|---|---|---|---|---|
| $S^{2-}$ | Silk fibroin-CuNCs | 422 (326) | 1.6% | 5-110 μM | 0.286 μM | (Zhang et al. 2019a) |
| Sulfide | CuNCs-CDs | 469, (365) | 6220.77% | 26-128 nM | 4.3 nM | (Wen et al. 2019) |
| TNT | PEI-CuNCs | 480 (355) | / | 0-8 nM | 14 pM | (Aparna et al. 2018) |
| Creatinine | GSH-CuNCs | 585 (360) | 27% ($Al^{3+}$) | 2.5-34 μg/L | 0.63 μg/L | (Jalili and Khataee 2018) |
| Dopamine | BSA-CuNCs | 406 (325) | / | 0.5-50 μM | 0.28 μM | (Miao et al. 2018) |
| Dopamine | BSA-CuNCs | 410 (355) | / | 0.1-0.6 nM | 0.024 nM | (Aparna et al. 2019b) |
| Dopamine | BSA-CuNCs | 440 (365) | / | 0.1-100 μM | 32 nM | (He et al. 2018) |
| Microcystin-LR | Hairpin DNA-CuNCs | 575 (340) | / | 0.005-1200 μg/L | 0.003 ng/L | (Zhang et al. 2019b) |
| Uracil-DNA Glycosylase | poly(T) DNA-CuNCs | 602 (400) | / | 0.1-10 U/L | 0.05 U/L | (Ling et al. 2019) |
| Folic acid | Ovalbumin-CuNCs | 625 (348) | 3.95% | 0.5-200 μM | 0.18 μM | (Li et al. 2019a) |
| Folic acid | Cys-CuNCs | 465 (365) | | 0.1-10 μM | 69.8 nM | (Han and Chen 2019) |
| Vitamin B1 | Ovalbumin-CuNCs | 560 (350) | 5.8% | 1-1000 nM | 380 pM | (Yang et al. 2018) |
| Doxycycline | Ovalbumin-CuNCs | 560 (350) | 5.8% | 1-1000 μM | 270 nM | (Yang et al. 2018) |
| Nitrite | Cys-CuNCs | 465 (365) | | 1-80 μM | 0.0954 μM | (Han and Chen 2019) |
| Bilirubin | HAS-CuNCs | 646 (390) | 3.6% | 1.25-7.5 μM and 5-145 nM and 28.75 μM | 35 nM | (Rajamanikandan and Ilanchelian 2019) |
| Rutin | BSA-CuNCs | 640 (320) | / | 0.1-100 μM | 0.02 μM | (Wang et al. 2018a) |
| Picric acid | GSH-CuNCs | 625 (410) | / | 9.9-43 μM | 2.74 μM | (Patel et al. 2018) |
| Picric acid | Ascorbic acid-CuNCs | 430 (350) | / | 2-40 μM | 0.98 μM | (Zhang et al. 2018c) |
| Picric acid | Cysteamine-CuNCs | 446 (365) | 2.3% | 1-80 μM | 139 nM | (Bao et al. 2018) |
| Picric acid | Cys-CuNCs | 494 (370) | / | 2.5-25 μM | 0.19 μM | (Shanmugaraj and John 2018) |

| Analyte | CuNCs | Em (Ex) nm | QY | Linear range | LOD | Ref. |
|---|---|---|---|---|---|---|
| Pyrophosphatase | TA-CuNCs | 438 (360) | / | 0.5-18 U/L | 0.19 U/L | (Liu et al. 2018a) |
| Pyrophosphatase | GSH-CuNCs | 615 (360) | / | 3-40 U/L | 1.3 U/L | (Ye et al. 2019) |
| Alkaline phosphatase | GSH-CuNCs | 610 (360) | / | 0.5-25 U/L | 0.15 U/L | (Geng et al. 2019) |
| Alkaline phosphatase | L-histidine-CuNCs | 410 (390) | | 0.5-40 U/L | 45 mU/L | (Hu et al. 2018a) |
| Matrix Metalloproteinase-7 | DNA-CuNCs | / | / | 0.01-100 ng/mL | 5.3 pg/mL | (Zhuang et al. 2019) |
| Water | AMTD-Ac-CuNCs | 398 (332) | 52.3% | / | / | (Cheng et al. 2018) |
| Protamine | BSA-CuNCs | 410 (330) | / | 3-12 ng/mL | 0.12 ng/mL | (Aparna et al. 2019a) |
| Heparin | BSA-CuNCs | 410 (330) | / | 6-9 ng/mL | 0.0406 ng/mL | (Aparna et al. 2019a) |
| Heparin | dBSA-CuNCs | 642 (350) | 2.32% | 1.25-250 ng/mL | 0.26 ng/mL | (Wu et al. 2018) |
| Nitrofurantoin | Adenosine-CuNCs | 417 (285) | / | 0.05-4 μM | 30 nM | (Wang et al. 2018b) |
| D-penicillamine | BSA-CuNCs | 400 (325) | / | 0.6-30 μg/mL | 0.54 μg/mL | (Ma et al. 2019) |
| M-dinitrobenzene | Cys-CuNCs | 492 (375) | / | 1.3-990 μM | 0.13 μM | (Hambarde et al. 2019) |
| Micrococcal nuclease | dsDNA-CuNCs | 570 (340) | / | 1-50 U/L | 1.0 U/L | (Qing et al. 2019a) |
| DNA adenine Methylation Methyltransferase | dsDNA-CuNCs | 590 (340) | / | 0.5-10 U/mL | 0.5 U/mL | (Gao et al. 2018) |
| Quinoline yellow | Cys-CuNCs | 494 (370) | / | 0.2-5.5 μM | 0.11 μM | (Shanmugaraj and John 2018) |
| Carbamazepine | CTAB-CuNCs | 415, 480 (290) | / | 0.2-20 μg/mL | 0.08 μg/mL | (Hatefi et al. 2019) |
| Glucose | GSH-CuNCs | 610 (365) | <0.5% | 0.1-2.0 mM | 32 μM | (Gou et al. 2019) |
| Histamine | TFTP-CuNCs | 590 (325) | / | 0.1-10 μM | 60 nM | (Han et al. 2018a) |
| Creatinine | BSA-CuNCs | 643 (525) | | 5-6 μM/L | 5 nM | (Rajamanikandan and Ilanchelian 2018) |

| Analyte | Probe | Em (Ex) | QY | Linear range | LOD | Reference |
|---|---|---|---|---|---|---|
| Trihexyphenidyl Hydrochloride | Cys-CuNCs–Ce(IV)/KMnO$_4$ system | / | | 0.1-10.0 μM | 49.0 nM | (Chen et al. 2018) |
| Iodine | TA-CuNCs | 460 (380) | | 20-100 μM | 18 nM | (Cao et al. 2018) |
| T4 polynucleotide Kinase phosphatase | dsDNA-CuNCs | 570 (340) | | 0.07-15 U/mL | 0.06 U/mL | (Zhang et al. 2018e) |
| Trinitrophenol | PVP-CuNCs | 518 (392) | 44.67% | / | 391 nM | (Li et al. 2019b) |
| Rifampicin | PEI-CuNCs | 492 (362) | 10.7% | 0-20 μM | 50 nM | (Tan et al. 2018) |
| Urea | Ascorbic acid-CuNCs | 498 (385) | 6.63% | 0.25-5 mM | 0.01 mM | (Deng et al. 2018b) |
| L-lysine | Ovalbumin-CuNCs | 440 (370) | / | 10.0 μM-1.05.5 mM | μM | (Zhang et al. 2018b) |
| Bilirubin | BSA-CuNCs | 404 (330) | / | 10-150 pM | 257 fM | (R et al. 2018) |
| Cytochrome c and trypsin | D-penicillamine-CuNCs | 645 (390) | / | 8-680 nM and 0.1-6.0 μg/mL | 0.83 nM and 20 ng/mL | (Hu et al. 2018b) |
| Choline | Lysozyme-CuNCs | 550 (440) | / | 0.1-80 μM | 25 nM | (Bu et al. 2018) |
| S1 nuclease | AT24-X6-hairpin-CuNCs | / | / | 5-80 U/L | 3 U/L | (Peng et al. 2018) |
| Streptavidin | ssDNA-CuNCs | 349 (617) | / | 0-200 nM | 0.47 nM | (Cao et al. 2017) |
| Biotin | ssDNA-CuNCs | 349 (617) | / | 10-1000 nM | 3.1 nM | (Cao et al. 2017) |
| DNA | dsDNA-CuNCs | 570 (340) | / | 0.01-10 nM | 7 pM | (Zhang et al. 2018d) |
| DNA | dsDNA-CuNCs | / | / | / | 1 lesion in 74 nucleotides. | (Singh et al. 2018) |
| MicroRNA-21 | dsDNA-CuNCs | 390 (525) | / | 100 aM-100 pM | 19.05 aM | (Liao et al. 2018b) |
| MicroRNA-155 | DNA-stabilized CuNCs | / | / | 100 aM-100 pM | 36 aM | (Zhou et al. 2018) |
| pH | L-proline-CuNCs | 458 (398) | / | 9-13 | / | (Han et al. 2018c) |
| pH | Cu$_9$S$_5$ NCs | 567 (406) | 6.8% | 3-10 | / | (Cheng et al. 2019) |

The symbol "/" indicated "not available", mainly due to the weak fluorescence signals.

## 4. CuNCs for bioimaging

Cancer cells generally show multiple differences in intracellular pH, temperature, enhanced receptor mediated, protein overexpression, and reactive oxygen/nitrogen production (Dutta et al. 2018). Therefore, the development of nanoprobes capable of monitoring these changes is particularly important. The preparation of nanoprobes for the selective detection of cancer cells remains a long-term challenge. In addition to specificity, the problems they face include their biological toxicity and complex biological environment. As of yet, CuNCs hold great promise in bioimaging applications because of their excellent biocompatibility. It is also worth mentioning that CuNCs with different fluorescence emissions could provide great opportunities in bioimaging. Compared to blue and green fluorescent metal nanoclusters, red fluorescent metal nanoclusters show significant advantages for *in vivo* imaging, e.g., high cell membrane penetration, and low tissue absorption (Zhang et al. 2014).

The controlled and highly luminescent CuNCs assemblies showed good stability in the cell imaging mentioned above (Zhou et al. 2019). Due to the PEG corona with good biocompatibility on the surface of CuNCs, the components demonstrated no obvious signs of toxicity to living cells. As displayed in Fig. 15a, CuNPs emitted bright red fluorescence in human epithelial cervical carcinoma cells (HeLa cells), and the corrected total cell fluorescence increased from 1 to 24 h of the incubation time. According to the experimental results, the synthesized CuNCs assemblies were not only resistant to complex physiological environments, but also unaffected by 1 M NaCl, 1 mM ascorbic acid, and 0.1 mM $H_2O_2$. In addition, the CuNCs assemblies were highly resistant to the adsorption of serum proteins and the effects of the harsh lysosomal acidic microenvironment. In Fig. 15b, red fluorescent CuNCs assemblies could colocalize their subcellular locations with green fluorescent protein (GFP)-labeled lysosomes, exerting great benefits on bioimaging applications.

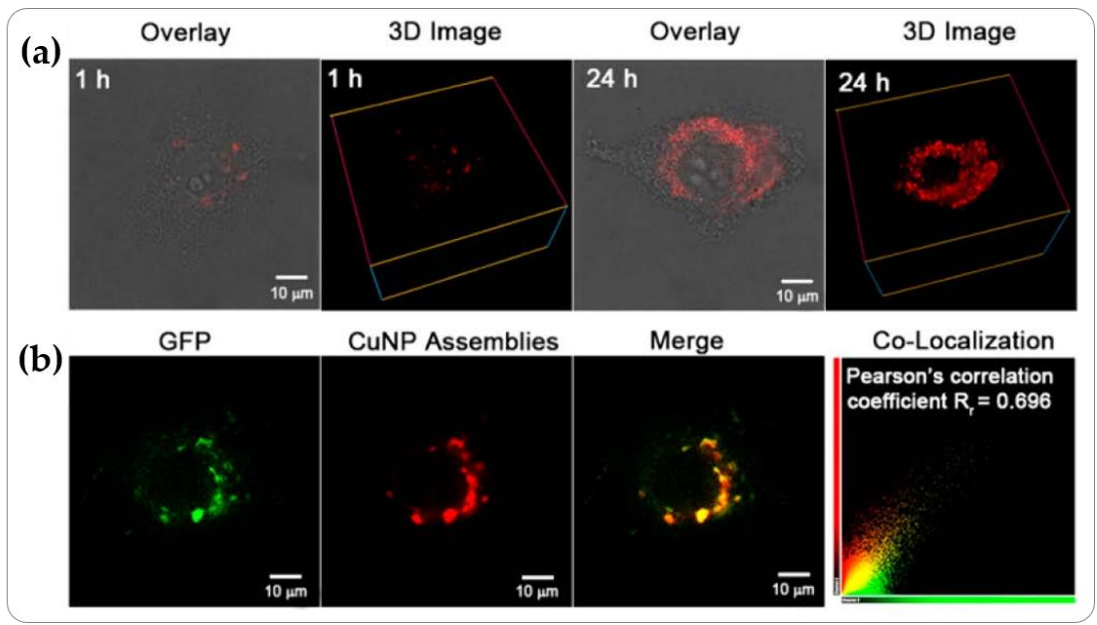

**Fig. 15.** (a) Overlay of bright and fluorescence field and three-dimensional (3D) images of the CuNPs assemblies in HeLa cells incubated at different time points. The corrected total cell fluorescence increased from 1 to 24 h of the incubation time. (b) Colocalization analysis between the CuNPs assemblies and GFP-labeled lysosomes (reprinted from (Zhou et al. 2019) with permission from American Chemical Society).

Xia et al. (Xia et al. 2018) reported that FA (folic acid)-CuNCs with high selectivity for folate receptor (FR) could be used for FR positive cell imaging. The prepared FA-CuNCs exhibited good photostability and water solubility. In many human cancer cells, FR is overexpressed on the surface, so the FA is usually used as a cell targeting agent because of the high affinity between FA and FR. In their study, HeLa cells and human lung carcinoma cells (A549 cells) were both selected. Since the number of surface-specific receptors in the two cancer cells were different, the intake of FA-CuNCs was also different. Taking advantage of this, FA-CuNCs were incubated with HeLa cells and A549 cells for 3 hours, respectively. Significant blue fluorescence could be observed in HeLa cells, and the average fluorescence intensity of HeLa cells showed a good linear relationship with the concentration of FA-CuNCs. However, almost no fluorescence was observed in A549 cells. In addition, FA-CuNCs had very low cytotoxicity. The survival rate of HeLa cells after incubation with 200 μg/mL FA-CuNCs for 20 hours was still greater than 80%. It was clear that FA functional groups in FA-CuNCs nanoclusters could significantly enhance the targeting ability and FA-CuNCs had great potential as fluorescent probes.

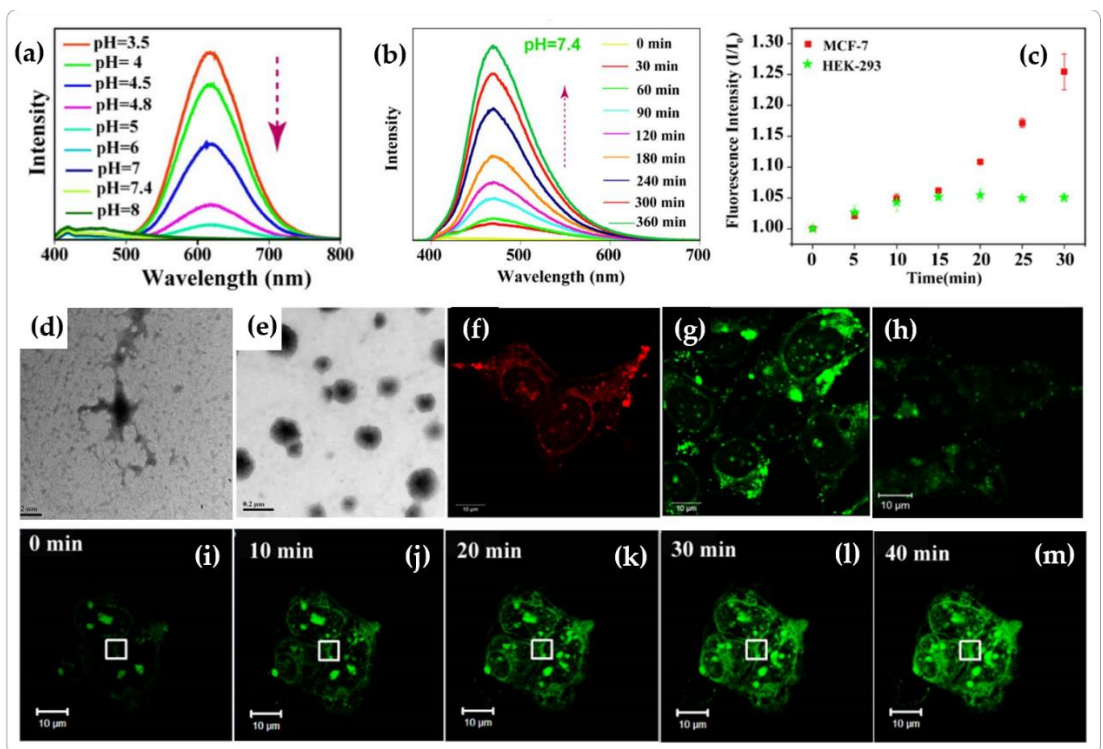

**Fig. 16.** (a) Fluorescence spectra of as-synthesized CuNCs at different pH values; (b) Time-dependent change in fluorescence spectra of CuNCs recorded at pH 7.4; (c) Change in the fluorescence intensity ratio ($I/I_0$) of the probe over time (min) recorded in MCF-7 and HEK-293 cells; TEM image of the CuNCs dispersion recorded (d) immediately after pH adjustment at 7.4 and (e) aging for 4h; (f) Fluorescence image of MCF-7 cells acquired at pH 4.5; Fluorescence image of (g) MCF-7 cells and (h) HEK-293 cells treated with the CuNCs probe for 1 h in pH 7.4 culture media; (i-m) Time-lapsed live cell confocal microscopy imaging of the MCF-7 cells treated with CuNCs at pH 7.4 (reprinted from (Dutta et al. 2018) with permission from American Chemical Society).

Dutta and colleagues (Dutta et al. 2018) successfully synthesized cysteine-stabilized CuNCs with good pH responsiveness (pH 4.5-8) in the presence of chitosan. In the study, CuNCs showed strong red fluorescence in aqueous solutions at a lower pH (4.5). As the pH of the solution increased, the fluorescence of CuNCs gradually disappeared (Fig. 16a). Interestingly, a unique phenomenon appeared at pH 7.4, where CuNCs exhibited a blue-green fluorescence emission that increased over time after aging (Fig. 16b). It could be clearly seen from the TEM image that, as the aging time increased, CuNCs gradually aggregated to form dense aggregates (Fig. 16d-e). The presence of chitosan enhanced the stability and biocompatibility of CuNCs. It not only ensured the AIEE effect, but also afforded the successful internalization of CuNCs in human breast adenocarcinoma cells (MCF-7 cells) and human embryonic kidney cells (HEK-293 cells). When MCF-7 cells were incubated with CuNCs, they showed a bright red fluorescence emission incubated in acidic medium (Fig. 16f). Similar to the case in an aqueous solution, further experiments under the same conditions at pH 7.4 showed that the MCF-7 cells, originally colorless, emitted bright green fluorescence over time. As shown in Fig. 16i-m, MCF-7 cells internalized by CuNCs exhibited surprisingly bright green fluorescence after aging for 40 min. In addition, a similar phenomenon was observed in HEK-293 cells. The difference was that the increase in emission observed in HEK-293 cells was relatively weaker compared to that observed in MCF-7 cells. If the rate of fluorescence growth in HEK-293 cells was considered to be linear, then the fluorescence intensity in MCF-7 cells increased exponentially (Fig. 16c), which can therefore be used to distinguish cell lines based on the differences in AIEE kinetic rates. Given the good pH responsiveness of CuNCs, cell health (cancerous or non-cancerous) could be determined by identifying changes in intracellular pH.

Recently, a dual emission ratio fluorescent probe was reported to be used to specifically identify $Ca^{2+}$ in a cell model (Liu et al. 2019b). The probe with good water solubility displayed two independent emission peaks at 590 nm and 690 nm. The intensity of the emission peak at 590 nm increased continuously as the concentration of $Ca^{2+}$ increased while the fluorescence intensity at 690 nm stayed constant (Fig. 17a). Based on its good biocompatibility, the developed fluorescent probe was used for biosensing and imaging of $Ca^{2+}$ in neurons. It was obvious that the green fluorescent channel became brighter as the concentration of $Ca^{2+}$ increased, while the red fluorescent channel remained unchanged (Fig. 17c). The fluorescent signal ratio ($F_{green}/F_{red}$) was used to quantify the amount of $Ca^{2+}$ in neurons. Interestingly, it was found that histamine could increase $Ca^{2+}$ levels in living cells. As shown in Fig. 17d, after the addition of histamine, the green fluorescence was gradually enhanced and kept constant at the 120s, while the red fluorescence intensity remained unchanged. Further experimentation demonstrated that the increase of $Ca^{2+}$ caused by histamine was different in certain parts of neurons: the concentration of $Ca^{2+}$ in the somatic and proximal axons was significantly higher than that in other parts. More importantly, the experiment revealed a possible mechanism of $O_2^{\cdot-}$ induced neuronal death. While exposed to 80 μM $O_2^{\cdot-}$, the $F_{green}/F_{red}$ value increased significantly. Based on the established calibration curve (Fig. 17b), the cytoplasmic $Ca^{2+}$ concentration increased to approximately 250 μM after the 600s, which exceeded normal concentrations. The excess of $Ca^{2+}$ may explain for neuronal death.

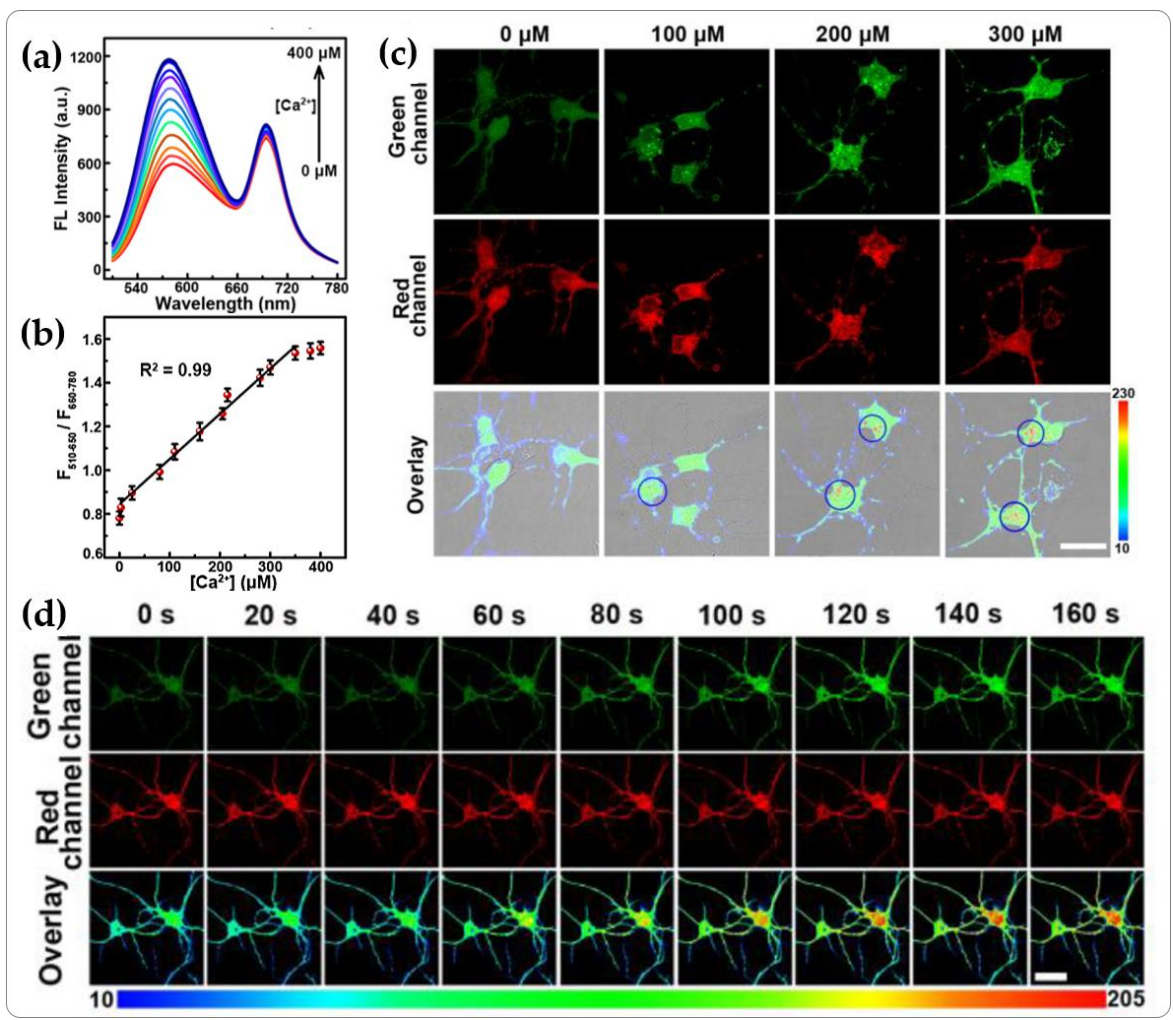

**Fig. 17.** (a) Fluorescence responses of CuNCs (90 μg/mL) with the addition of various concentrations of $Ca^{2+}$ (0, 2, 25, 80, 110, 160, 205, 215, 280, 300, 350, 380, 400 μM) in cell lysis buffer. (b) Calibration curve between $F_{510-650}/F_{660-780}$ and various concentrations of $Ca^{2+}$. Data was obtained from Fig. 17a. (c) Confocal fluorescence microscopy images of neurons collected from different channels after the neurons were coincubated with CuNCs probe (90 μg/mL) in the presence of different concentrations of $Ca^{2+}$ (0, 100, 200, and 300 μM), respectively. Bright areas within the blue circle indicate uneven distribution of cytosolic $Ca^{2+}$. (d) Time-tracking of confocal fluorescence microscopic images of neurons was stimulated by 50 μM histamine for different times (reprinted from (Liu et al. 2019b) with permission from American Chemical Society).

## 5. Conclusion and Perspectives

In this review, we focused on the fluorescence of CuNCs. The factors affecting the CuNCs fluorescence are primarily divided into four categories: (1) size-dependent fluorescence; (2) Cu⋯Cu interaction; (3) the solvent effect; and (4) aggregation-induced luminescence. This review highlights the effects of aggregation-induced enhanced emission and its importance in detection processes. The unique fluorescent properties of CuNCs provide great opportunities in sensing and bioimaging applications. Many detection methods have been used for actual testing and have achieved satisfactory performances. According to the reported results, there are a variety of CuNC nanoprobes for certain substrates. However, the detection environment (pH, concentration range, etc.) for each nanoprobe is different. Therefore, we should specifically choose optimized CuNCs according to the environment in which the substrate is placed. Furthermore, it is worth mentioning that nanometer-sized CuNCs also have achieved good results in catalytic performance.

From current research progress, most CuNC nanoprobes have been used for environmental testing, as well as *in vitro* testing. More meaningful, *in vivo* probes are seldomly reported. As the ultimate goal of biological probe usage, biological probes are subject to a variety of factors in their specific applications. Although CuNCs have good biocompatibility, the stability and selectivity of probes are greatly challenged by *in vivo* complex environments, such as pH change, ascorbic acid, $H_2O_2$, and even lysosomes. Although significant advances having been made in the functionalization and biomedical applications of fluorescent CuNCs over the past few years, there are still challenges that need to be addressed in the future. Compared with AuNCs and AgNCs, CuNCs are more likely to oxidize and thus have lower quantum yields. Overall, improvement in fluorescence emissions and further enhancement of quantum yield remain the focus of future developments. Furthermore, although the AIEE effect can significantly improve the intensity of CuNCs fluorescence, the stability of the aggregates induced by this strategy is somewhat challenged by biological environments. Additionally, the use of organic solvents limits its bio-related applications. The pH-dependent AIEE effect may be the entry point for study *in vivo*. Presently, another problem faced in the design of clusters is to determine the key factors of quantum yield and emission wavelength of CuNCs. The clear image underlying the fluorescence mechanism, precise determination of crystal structure, including the arrangement of metal atoms and metal-ligand bonding, will be an important step forward in the development of CuNCs.

**Acknowledgments:** B.C. acknowledges financial support from the National Natural Science Foundation of China (51803161 and 51533007). J.C. acknowledges the University of California, Los Angeles for the startup support.

**Declaration of competing interest:** The authors declare that they have no competing financial interests.